\documentclass[10pt,conference]{IEEEtran}
\IEEEoverridecommandlockouts
\usepackage{cite}
\usepackage{amsmath,amssymb,amsfonts}
\usepackage{algorithmic}
\usepackage{subfig}
\usepackage{graphicx}
\usepackage{multirow}
\usepackage{makecell}
\usepackage{multicol}
\usepackage{hyperref}
\usepackage{colortbl}
\definecolor{mygray}{gray}{.9}
\usepackage{textcomp}
\usepackage{url}
\usepackage{xcolor}
\usepackage{tcolorbox}
\usepackage{xspace}
\def\BibTeX{{\rm B\kern-.05em{\sc i\kern-.025em b}\kern-.08em
    T\kern-.1667em\lower.7ex\hbox{E}\kern-.125emX}}
\begin{document}

\newcommand{\zqj}[1]{{\color{blue}{\textbf{[ZQJ: #1]}}}}
\newcommand{\dao}[1]{{\footnotesize{\textcolor{blue}{[Dao: {#1}]}}}\xspace}

% 第一,可行性feasibility  题目修改--pre-trained model还是large llm
% 第二，期刊--开源大模型/biyuan大模型--codellama---chatgpt
% 第三，增加一个rq，深入探讨PTM在识别错误的样本上的原因
% 第四，增加两个讨论，探讨PTMs早不同警告型上的预测情况，以及在闭源大模型chatgpt上的概况
% 第五，不同的大模型上，版本size对结果的影响

\title{Pre-trained Model-based Actionable Warning Identification: A Feasibility Study
% Pre-trained Model-based Actionable Warning Identification: How Far are We?\\
% {\footnotesize \textsuperscript{*}Note: Sub-titles are not captured in Xplore and should not be used}
}

\author{Anonymous Author(s)}

\author{
\IEEEauthorblockN{Xiuting Ge$^{1,2}$, Chunrong Fang$^{1,*}$, Quanjun Zhang$^{1}$, Daoyuan Wu$^{2}$, Bowen Yu$^{1}$, Qirui Zheng$^{1}$, An Guo$^{1}$, \\ Shangwei Lin$^{2}$, Zhihong Zhao$^{1}$, Yang Liu$^{2}$, Zhenyu Chen$^{1}$}
\IEEEauthorblockA{
\textit{$^1$ The State Key Laboratory for Novel Software Technology, Nanjing University, China} \\
\textit{$^2$ School of Computer Science and Engineering, Nanyang Technological University, Singapore} \\
\textit{$^{*}$ Corresponding author} \\
dg20320002@smail.nju.edu.cn, fangchunrong@nju.edu.cn, quanjun.zhang@smail.nju.edu.cn, \\ daoyuan.wu@ntu.edu.sg, \{201250070, 201250229, guoan218\}@smail.nju.edu.cn, shang-wei.lin@ntu.edu.sg, \\zhaozhih@nju.edu.cn, yangliu@ntu.edu.sg, zychen@nju.edu.cn
}
}

% guoan218@smail.nju.edu.cn

% dg20320002@smail.nju.edu.cn, fangchunrong@nju.edu.cn, \{quanjun.zhang, 201250070, 201250229\}@smail.nju.edu.cn, {daoyuan.wu, shang-wei.lin}@ntu.edu.sg, zhaozhih@nju.edu.cn, yangliu@ntu.edu.sg, zychen@nju.edu.cn

\maketitle

\begin{abstract}
% With the advances in Machine Learning (ML), various approaches have been proposed to incorporate ML techniques into AWI. 
% However, the performance of these ML-based AWI approaches still remains restricted due to the reliance on a limited number of labeled warnings to develop the AWI classifier. 
% which are self-supervisedly trained on large-scale unlabeled corpora and fine-tuned on a limited labeled data to be adapted for downstream tasks,

Actionable Warning Identification (AWI) plays a pivotal role in improving the usability of static code analyzers. 
Currently, Machine Learning (ML)-based AWI approaches, which mainly learn an AWI classifier from labeled warnings, are notably common. 
However, these approaches still face the problem of restricted performance due to the direct reliance on a limited number of labeled warnings to develop a classifier. 
Very recently, Pre-Trained Models (PTMs), which have been trained through billions of text/code tokens and demonstrated substantial success applications on various code-related tasks, could potentially circumvent the above problem. 
Nevertheless, the performance of PTMs on AWI has not been systematically investigated, leaving a gap in understanding their pros and cons. 
In this paper, we are the first to explore the feasibility of applying various PTMs for AWI. 
By conducting the extensive evaluation on 10K+ SpotBugs warnings from 10 large-scale and open-source projects, we observe that all studied PTMs are consistently 9.85\%$\sim$21.12\% better than the state-of-the-art ML-based AWI approaches. 
Besides, we investigate the impact of three primary aspects (i.e., data preprocessing, model training, and model prediction) in the typical PTM-based AWI workflow. 
Further, we identify the reasons for current PTMs' underperformance on AWI.
Based on our findings, we provide several practical guidelines to enhance PTM-based AWI in future work. 
\end{abstract}

% Further, we analyze the reasons for PTMs' underperformance on AWI. the limitations of PTM on AWI.

% The experimental results  describe that all studied PTMs are consistently 11.50\%$\sim$16.85\% better than the state-of-the-art ML-based AWI approaches. 
% Besides, we investigate the impact of three primary components (i.e., data preprocessing, model training, and model prediction) in the Typical PTM-based AWI workflow. 
% Moreover, based on the above findings, we provide several practical guidelines to further enhance AWI in future work. 

% we conduct an additional discussion about the preliminary application of ChatGPT on AWI, thereby signifying the capacity and limitation of PTMs on AWI.

\section{Introduction} \label{intro}
Static Code Analyzers (SCAs) can automatically scan software codebases and reveal potential defects without executing the program \cite{static}. 
Despite the benefits of SCAs in software defect detection \cite{useful3, useful5}, SCAs are still underused in practice due to reporting an overwhelming number of unactionable warnings, especially false positives \cite{rebalance, featureselection, shab}.  
Manually identifying warnings into actionable and unactionable ones is time-consuming and error-prone \cite{datasetheihei}.
As such, the tremendous unactionable warnings and the tedious manual inspection cost pose significant barriers to the usability of SCAs.

To alleviate the above problem, different approaches \cite{andreasen2017systematic} have been proposed to optimize the precision of SCAs from the vendor's perspective, thereby reducing the number of false positives reported by SCAs. 
However, since the trade-off between precision and recall is non-trivial in the static analysis \cite{rice}, it is inevitable for SCAs to report false positives. 
Despite retaining an initially high precision, SCAs could undergo a decline in defect detection performance as the nature of defects changes over time \cite{S44}. 
The continuous maintenance and update to make SCAs overcome the concept drift could be an expensive endeavor \cite{bielik2017learning}.
As such, an alternative approach \cite{survey1}, i.e., \textit{Actionable Warning Identification (AWI)}, has been proposed to postprocess warnings reported by SCAs from the user’s perspective, thereby identifying actionable warnings from all reported warnings.
Unlike the existing precision optimization approaches \cite{andreasen2017systematic} that refine the complex static analysis techniques before the usage of SCAs, AWI focuses on leveraging various postprocessing techniques (e.g., clustering, ranking, pruning, or simplifying manual inspection) to classify or prioritize warnings after the usage of SCAs.
It indicates that AWI is independent of specific SCAs with different static analysis techniques.
More importantly, AWI can be equipped with various postprocessing techniques to augment SCAs, where these postprocessing techniques complement the capabilities of SCAs to report more precise results.

In the existing AWI approaches, Machine Learning (ML)-based AWI approaches are notably popular due to ML's strong ability to learn subtle and previously unseen patterns from historical data. 
The general process of ML-based AWI approaches is to utilize ML models to train the AWI classifier from labeled warnings and use this classifier to identify actionable warnings from unlabeled ones \cite{ge2023machine}. 
However, the effectiveness of these approaches is still limited because the AWI classifier is generally established on a small number of labeled warnings \cite{empirical, empiricalex, sanxing, MAPL, smartcontract, S44}. 
It indicates that the true power provided by ML techniques has not been fully unleashed on AWI.
% In particular, there only involve 400 warnings in the work of Koc et al. \cite{empirical}.

% The recently rapid developments in Pre-Trained Models (PTMs) provide an alternative solution for existing ML-based AWI approaches because PTMs 
The rapid development of ML techniques has spurred the emergence of Pre-Trained Models (PTMs). 
Different from the supervised learning of ML-based AWI approaches on labeled warnings, PTMs are trained in a self-supervised fashion based on the tremendous unlabeled corpora and can be used for downstream tasks by fine-tuning limited labeled samples \cite{pretrainedmodel}. 
Currently, PTMs have exhibited remarkable performance in a variety of code-related tasks (e.g., software vulnerability repair \cite{zhang2023pre}). 
To alleviate the problem of existing ML-based AWI approaches, the unique characteristics and recent breakthroughs of PTMs inspire us to apply PTMs for AWI \cite{S44}. 
However, the literature does not systematically investigate the actual power of modern PTMs on AWI, thereby failing to understand the pros and cons of PTM-based AWI.

% including (1) the effectiveness of PTM-based AWI, (2) the impact of different data preprocessing ways (i.e., warning context construction and abstraction) on PTM-based AWI, (3) the impact of different model training components (i.e., pre-training and fine-tuning component) on PTM-based AWI, and (4) the performance discrepancy of PTM-based AWI in different model prediction scenarios (i.e., with-in and cross project AWI).
% declines the AWI performance in three out of five PTMs because the warning abstraction may hinder the utilization of generic knowledge in the PTMs for AWI

% (1) the effectiveness of PTM-based AWI, (2) the impact of different data preprocessing ways (i.e., warning context construction and abstraction) on PTM-based AWI, (3) the impact of different model training components (i.e., pre-training and fine-tuning) on PTM-based AWI, (4) the performance discrepancy of PTM-based AWI in different model prediction scenarios (i.e., with-in and cross project AWI), and (5) the of PTMs on AWI.
To bridge the above gap, we perform the first extensive study to explore the feasibility of PTMs on AWI.
We first investigate the effectiveness of PTM-based AWI. Based on the typical PTM-based AWI workflow, we analyze the impact of the data preprocessing ways, model training components, and model prediction scenarios. 
Further, we identify the underperformance of current PTMs on AWI.
By conducting experiments on more than 10K+ SpotBugs \cite{spotbigs} warnings from 10 large-scale and real-world Java projects and five representative PTMs with encode-only, encoder-decoder, and decoder-only architectures, the results demonstrate that (1) five PTMs on AWI achieve the AUC of 62.70\%$\sim$70.77\%, which outperform the State-Of-The-Art (SOTA) ML-based AWI approach by 9.85\%$\sim$21.12\%; (2) in the data preprocessing, the warning context from the method containing a warning can be consistently better than that from the warning line numbers. Surprisingly, the warning context abstraction does not necessarily improve the performance of PTMs on AWI as the abstraction operation could hinder the utilization of PTM's generic knowledge on AWI; (3) the pre-training and fine-tuning components in the model training are beneficial for PTM-based AWI; (4) in the model prediction, PTMs can achieve better performance in the within project AWI scenario than in the cross project AWI scenario; and (5) PTMs struggle on tasks involving similar or even the same contexts in actionable and unactionable warnings, insufficient or unavailable warning contexts, and lacking adequate warnings with diverse types in the training set.
Based on the above findings, we highlight practical guidelines (e.g., the warning context refinement) for the future PTM-based AWI field.
% In addition, we provide the discussion to investigate the performance of ChatGPT on AWI. The preliminary results demonstrates that -----.
% PTMs underperform in classifying the warnings with similar and insufficient contexts. Also, when lacking sufficient and diverse warnings for the fine-tuning, PTMs cannot perform well on AWI.

In summary, this paper makes the following contributions. 
\begin{itemize}
    \item \textbf{New perspective.} We incorporate recent advances of PTM into AWI community. Besides, we conduct a systematic evaluation to unveil the substantial improvement of PTM on AWI. 
    We believe that our study yields the best of current ML and static analysis fields, i.e., ML augments the usability of existing SCAs.
    % i.e., PTM can augment the usability of SCAs to acquire more precise results.
    % We incorporate the PTMs into AWI and conduct an extensive evaluation, demonstrating that the PTMs provide substantial improvements on AWI. 
    % By leveraging the PTMs to augment existing SCAs, our study yields the best of both fields, where the PTMs supplement the capabilities of current SCAs.
    % We incorporate PTMs into AWI. Besides, an extensive evaluation demonstrates that PTMs provide substantial improvements on AWI.   a crucial software engineering problem
    % we bride the gap
    % We believe that by incorporating PTMs into AWI, our study yields the best of ML and static analysis fields, where PTMs augment the usability of SCAs to generate more precise results.
    % We believe that by leveraging PTMs for AWI, our study yields the best of ML and static analysis fields, where PTMs augment the usability of SCAs to generate more precise results.     
    % and an addition discussion about the performance of ChatGPT on AWI.

% the application of PTMs for AWI in future research, and 
    % such as en the benefits of pre-training and fine-tuning in PTMs

    \item \textbf{Extensive study.} We are the first to conduct an extensive study to explore the feasibility of PTMs on AWI, including a detailed comparison between SOTA ML-based and PTM-based AWI approaches, a thorough investigation about the impact of primary aspects (i.e., data preprocessing, model training, and model prediction) in the typical PTM-based AWI workflow, and an in-depth analysis about the challenges of PTMs on AWI.

    \item \textbf{Practical guidelines.} We highlight several practical guidelines in future PTM-based AWI research, e.g., the warning context refinement to further enhance AWI. 

    \item \textbf{Available artifacts.} We release the studied warning dataset and the experimental scripts in a public repository \cite{github} for replication and future research. 
    
    % share the studied warning dataset and the experimental scripts in a public repository \cite{github} for replication and future research. 
    
\end{itemize}

\section{Background and Related Work} \label{background}

\subsection{Static Analysis Warnings} \label{saw} 
SCAs can detect various defects in the codebase, e.g., security issues and code smells. 
The existing AWI studies \cite{ViolationTracker, mining, golden, shab} denote such defects as static analysis warnings, alerts, alarms, or violations. 
In our study, such defects are simply denoted as warnings.
To help developers quickly locate and understand defects, each warning is generally equipped with category, priority, message, and location. 
Of these, the location often consists of the class and method information containing a warning as well as the warning line numbers. 
% To clearly describe a warning, Fig. \ref{fig:example} shows a warning example obtained from SpotBugs \cite{spotbigs}. 
% \begin{figure}
%     \centering
%     \includegraphics[scale=0.155]{figure/example.jpg}
%     \caption{An example of the warning reported by SpotBugs. \dao{This fig is not clear, especially the light blue and green color; the four sentences at the right may use the dark red color; also add a little space between lines}}
%     \label{fig:example}
% \end{figure}

Based on whether warnings are acted on and fixed by developers, warnings can be divided into actionable and unactionable ones \cite{mining, ViolationTracker, golden, shab}. 
An actionable warning, including a true defect or a warning concerned by SCA users, is acted on and fixed by developers via the warning-related source code changes.
Conversely, an unactionable warning might be a false positive warning due to the inherent problem (i.e., over-approximation) of SCAs \cite{rice}, an unimportant warning for SCA users, or just an incorrectly reported warning due to limitations of SCAs \cite{limit}. Thus, an unactionable warning is not acted on or fixed by developers. 

Formally, given a set of commits \emph{$C = \{c_1, ..., c_i, ..., c_n\}$} in a project (\emph{n} is the latest commit), a SCA is used to scan the source code of \emph{$c_i$} and a set of warnings \emph{$W_i = \{w_{i1}, ..., w_{ij}, ..., w_{im}\}$} (\emph{m} is the number of warnings in \emph{$c_i$}) is obtained. 
If \emph{$w_{ij}$} disappears via the warning-related source code change in any commit from \emph{$c_{i+1}$} to \emph{$c_n$}, \emph{$w_{ij}$} is denoted as \textbf{an actionable warning}. 
If \emph{$w_{ij}$} persists from \emph{$c_{i+1}$} to \emph{$c_n$}, \emph{$w_{ij}$} is denoted as \textbf{an unactionable warning}.

\subsection{ML-based AWI approaches} \label{awi} 
In general, ML-based AWI approaches first extract features from the warning report or the warning-related source code, learn an AWI classifier from these extracted features of labeled warnings via traditional ML models or Deep Learning (DL) models, and utilize this classifier to identify unlabeled warnings into actionable and unactionable ones.
Based on the classification of adopted ML models, ML-based AWI approaches can be roughly divided into traditional ML-based and DL-based AWI ones \cite{ge2023machine}. 
Benefiting from DL techniques' powerful feature representation ability, DL-based AWI approaches generally outperform traditional ML-based AWI approaches \cite{empirical, empiricalex, S44}. 
However, an effective AWI classifier extremely relies on labeled warnings and there is a limited number of labeled warnings. As such, the performance of existing ML-based AWI approaches is still restricted.

\subsection{Pre-trained Models} \label{ptm}
% Nowadays, the pre-trained models have been exhibited tremendous capability across a wide range of code-related tasks, such as the text classification \cite{liu2021finbert} and software vulnerability repair \cite{zhang2023pre}.

% the pre-trained models undergo a initial stage of self-supervised training on large-scale and unlabeled corpora to distill the generic representation and subsequently employ such the generic representation to handle the downstream tasks by fine-tuning a limited labeled corpus
% perform the self-supervised training
PTMs pre-train transformer-based models on large-scale and unlabeled corpora to distill the generic representation and then employ such generic representation to handle downstream tasks by fine-tuning a limited number of labeled corpus \cite{pretrainedmodel}. 
According to different architectures, PTMs can be classified into encoder-only, decoder-only, and encoder-decoder models \cite{llmclassification}. 
The encoder-only model, focusing on solely transforming the input data into the latent representation, is good at understanding tasks like text classification. 
The decoder-only model, aiming to decode output sequences from a given representation of the input data, is good at generating tasks like text completion. 
The encoder-decoder model combines both an encoder and a decoder into a single architecture, which is capable of handling sequence-to-sequence tasks. 

In the PTM-based AWI field, given a targeted warning with the associated context \emph{$X = \{x_1, ..., x_k\}$}, \emph{$x_k$} is the \emph{$k_{th}$} code token in the warning context. 
Taking \emph{$X$} as the input, PTM-based AWI relies on \emph{Pr($X$;$\theta$)} to output a class label \emph{$y$}. 
The weight $\theta$ is obtained from the transformer that makes up the encoder and decoder. \emph{$y$} is a binary value, where \emph{y = 0/1} denotes an unactionable/actionable warning respectively.

\subsection{Related work}
Similar to our study, Kharkar et al. \cite{S44} attempt two transformer-based models for AWI based on the warning context. One is to learn an AWI classifier from labeled warnings via CodeBERTa, while the other is to generate the warning-related code recommendation to infer the legality of warnings via GPT-C.
However, our study is different from their work in three aspects.
First, instead of using GPT-C for code completion recommendation in AWI, our study considers AWI to be a code classification task. Such a concept designn operation alignsis in line with the naturalness of AWI, thereby better understanding the outputs of PTM-based AWI.
Second, their work only adopts two PTMs with encoder-only and decoder-only models for AWI. By contrast, our study elaborately selects five PTMs for AWI, which span three typical PTM architecture categories.
Third, compared to their work, our study conducts a more thorough exploration of PTM-based AWI. That is, our study follows the typical PTM-based AWI workflow to investigate the impact of different aspects in the data preprocessing, model training, and model prediction stages.

\begin{figure}
    \centering
    \includegraphics[scale=0.51]{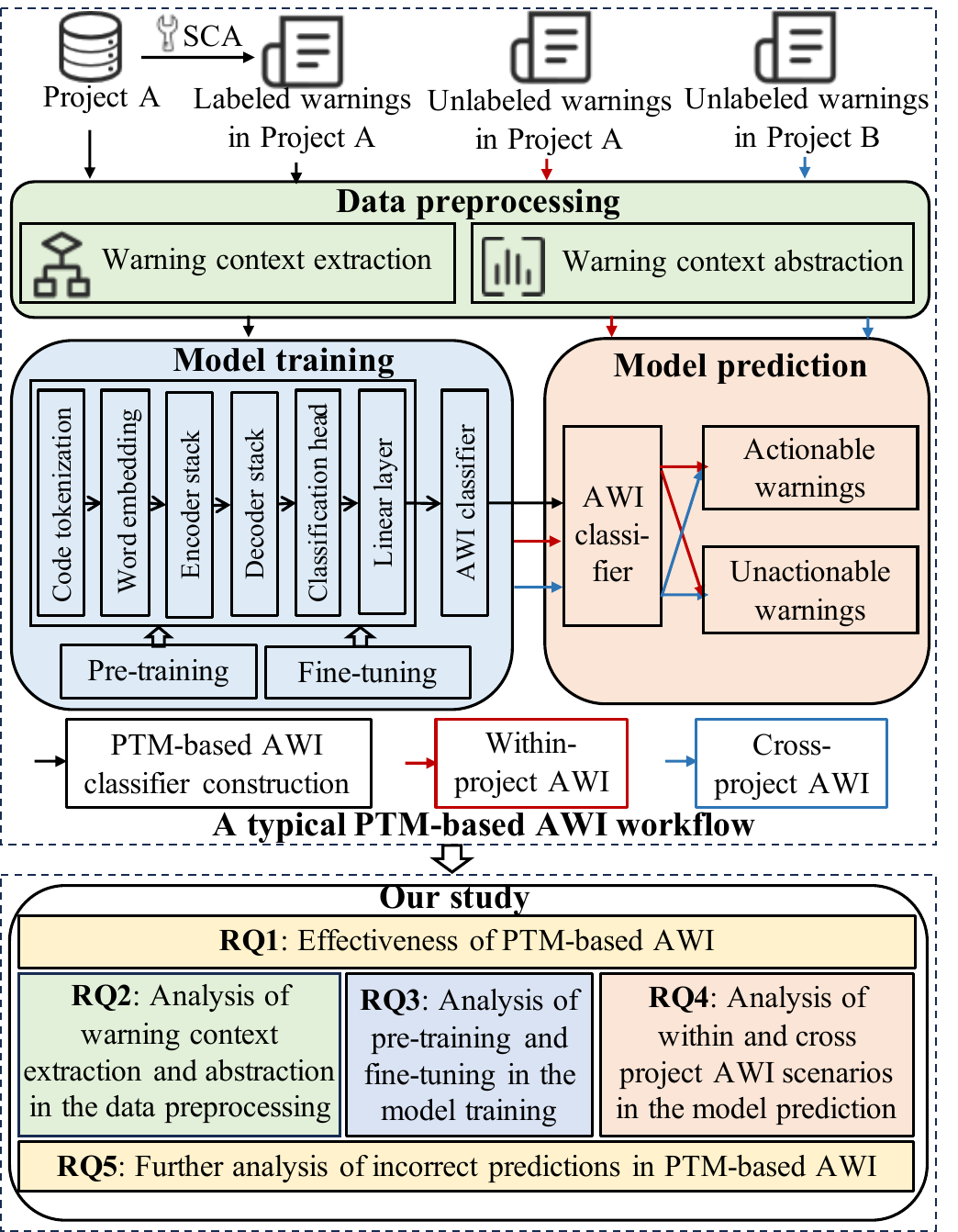}
    \caption{Overview of our study. 
    }
    \label{fig:overview}
\end{figure}
\section{Study Overview} \label{approach}

\subsection{Typical PTM-based AWI Workflow} \label{overview}
Fig. \ref{fig:overview} shows a typical PTM-based AWI workflow with data preprocessing, model training, and model prediction stages. 

% The constructed warning context can help reason or understand whether a warning is actionable.
% The abstracted warning context can significantly reduce the size of vocabulary, thereby facilitating improving the generalizability of the AWI classifier \cite{baselinednn2}. 
\textbf{Data preprocessing.} Given a warning reported by a SCA as input, the processed context of this warning is returned. According to the existing ML-based AWI studies \cite{empirical, empiricalex, smartcontract, MAPL, sanxing, S44}, the data preprocessing stage mainly involves the warning context extraction and abstraction. 
The warning context extraction acquires the warning-related source code based on the warning information. 
The warning context abstraction renames some special words (e.g., identifiers and literals) in the warning context to a pool of predefined code tokens. 
% After performing the tokenisation on the warning context, a sequence of tokens for the warning is obtained. 

% Unlike the generation task that utilizes this last hidden state to generate the output sequence, PTM-based AWI, which is a classification task in our study, then uses the classification head to reduce the dimensionality of this last hidden state and finally employs a softmax layer to output a binary value of this targeted warning.
% Our study systematically applies PTM for AWI. 

% A PTM-based AWI classifier is obtained when the processed contexts of labeled warnings pass through the code tokenization, word embedding, encoder stack, decoder stack, classification head, and softmax layer. 
% An AWI classifier is output when PTM undergoes the model training on the processed contexts of labeled warnings. 
\textbf{Model training.} A PTM-based AWI classifier is first established on the top of the transformer \cite{vaswani2017attention} and the mapping from warning context to warning label is optimized by updating the parameters of the designed classifier.
Similar to the vanilla transformer architecture \cite{vaswani2017attention}, PTMs often initiate with an encoder stack and a decoder stack, and culminate with a linear layer equipped with softmax activation.
In AWI, taking the processed warning context as input, PTM first splits the input into words via code tokenization. 
Second, PTM performs the word embedding to yield the representation vectors for the tokenized warning context.
Third, PTM feeds these vectors into the encoder and decoder stacks to output a last hidden state. 
Fourth, PTM adds a classification head for this last hidden state to obtain the logits. 
Fifth, PTM employs a linear layer with softmax activation for the obtained logits to acquire the probability distribution of binary warning labels. 

Generally, PTM involves two essential components, i.e., pre-training and fine-tuning \cite{pretrainedmodel}. 
In AWI, the pre-training component is related to whether a PTM-based AWI classifier is obtained by pre-training over large-scale programming language corpora. In contrast, the fine-tuning component is related to whether a PTM-based AWI classifier is obtained by fine-tuning on a limited number of labeled warnings.

% Unlike the generation task that utilizes this last hidden state to directly look up the output sequence, PTM-based AWI is a classification task in our study.

% to reduce the dimensionality of this last hidden state and employs a softmax layer for the handled state 
% output a binary value for this warning.

% usually starts with an encoder stack and a decoder stack and ends with a linear layer with softmax activation. 
% In the context of AWI, the pre-trained model first utilize the word embedding technique to transform the tokenized warning context into vectors. Then, the pre-trained model feeds these vectors into the encoder stack to derive the hidden state and further puts the hidden state into the decoder stack. 
% Finally, the output of the decoder stack is passed into a linear layer with softmax activation, thereby generating the probability distribution that represent a warning label.
% \zqj{it seems that classification do not use a decoder; however, I'm not sure how it works for deocder-only models, such as codegpt}

% \dao{Model prediction?}
\textbf{Model prediction.} The well-trained PTM-based AWI classifier is used to classify unlabeled warnings into actionable and unactionable ones during the model prediction. 
Based on different project sources between labeled warnings in the model training and unlabeled warnings in the model prediction, there are within and cross project AWI scenarios. 
When labeled warnings used for the model training and unlabeled warnings used for the model prediction are from the same project, it is called the within project AWI scenario. 
By contrast, when labeled warnings used for the model training and unlabeled warnings used for the model prediction are from different projects, it is called the cross project AWI scenario.

% \dao{Any dedicated methodology we need to do for customizing PTM to the context of AWI?}
% \dao{Because Fig. 2 looks like there are many contents, but here the description is limited.}
\begin{table*}[]
    \centering
    \begin{tabular}{|c|c|c|c|c| c|c|c|c| c|c|}
    % \begin{tabular}{ccccccccc}
    \hline
        \textbf{No.} & \textbf{Project}		&	\textbf{Time period}	&	\textbf{\#LoC}	&	\textbf{\#Commits}	& \textbf{\#C. commits}	&	\textbf{\#UW}	&	\textbf{\#AW}	&	\textbf{\#W}	& \textbf{\#Category} & \textbf{\#Type}\\	\hline \hline
1 &bcel	&	2001/10/29 $\sim$ 2023/02/11	&	10k+$\sim$168k+	&	2400	&	1913	&	595	&	30	&	625 & 7 & 41	\\	\hline
2 & codec		&	2003/04/26$\sim$2022/11/26	&	5k+$\sim$55k+	&	2296	&	1966	&	595	&	70	&	665 & 6 & 31	\\	\hline
3 & collections		&	2001/04/14$\sim$2022/11/02	&	1k+$\sim$136k+	&	3810	&	1144	&	642	&	3	&	645 & 6 & 29	\\	\hline
4 &configuration	&	2003/12/23 $\sim$ 2022/12/24	&	20k+$\sim$134k+	&	3743	&	3169	&	2843	&	62	&	2905	&10 & 56\\	\hline
5 &dbcp		&	2001/04/15$\sim$2023/02/10	&	8k+$\sim$55k+	&	2791	&	638	&	150	&	15	&	165	& 9 & 33 \\	\hline
6 &digester	&	2001/05/03 $\sim$ 2023/02/04 	&	3k+$\sim$54k+	&	2233	&	1622	&	620	&	30	&	650	& 9 & 40\\	\hline
7 &fileupload	&	2002/03/24$\sim$ 2022/10/25	&	2k+$\sim$16k+	&	1284	&	1064	&	528	&	40	&	568	& 6 & 26 \\	\hline
8 &mavendp		&	2006/04/10$\sim$2022/10/30	&	5k+$\sim$37k+	&	1165	&	748	&	900	&	18	&	918	& 7 & 37 \\	\hline
9 &net	&	2002/04/03$\sim$2022/11/08	&	50k+$\sim$57k+	&	2683	&	1672	&	687	&	31	&	718	& 8 & 50 \\	\hline
10 &pool	&	2001/04/15$\sim$2023/02/10 	&	6k+$\sim$34k+	&	2656	&	1638	&	2106	&	175	&	2281	& 8 & 39\\	\hline \hline
Sum & /	&	/	&	110k+$\sim$746k+	&	25061	&	15574	&	9666	&	474	&	10140	& 10 & 137 \\	\hline
    \end{tabular}
    \caption{Dataset information. \#C. commits, \#UW, \#AW, \#W, \#Category, and \#type are the number of commits compilabled by SpotBugs, unactionable warnings, actionable warnings, all warnings, distinct categories, and distinct types respectively.}
    \label{tab:dataset}
\end{table*}

\subsection{Research Questions}
Inspired by the typical PTM-based AWI workflow, we investigate the following Research Questions (RQs).

\textbf{RQ1}: How is the performance of PTMs on AWI in comparison to the SOTA ML-based AWI approach?

\textbf{RQ2}: How do the data preprocessing ways affect the performance of PTMs on AWI?
\begin{itemize}
    \item \textbf{RQ2.1}: What is the impact of warning context extraction?
    \item \textbf{RQ2.2}: What is the impact of warning context abstraction?
\end{itemize}

\textbf{RQ3}: How do the model training components affect the performance of PTMs on AWI?
\begin{itemize}
    \item \textbf{RQ3.1}: What is the role of a pre-training component?
    \item \textbf{RQ3.2}: What is the role of a fine-tuning component?
\end{itemize}

\textbf{RQ4}: How do the model prediction scenarios affect the performance of PTMs on AWI?

\textbf{RQ5}: What are root causes of incorrect predictions in the current PTM-based AWI approach?

% How is the performance difference of PTMs on AWI in the model prediction scenarios?

% How do the model prediction scenarios affect the performance of PTMs on AWI?

% \begin{itemize}
%     \item \textbf{RQ4.1}: What is the performance of PTMs in the within project AWI scenario?
%     \item \textbf{RQ4.2}: What is the performance of PTMs in the cross project AWI scenario?
% \end{itemize}

\subsection{Evaluation Metrics} 
We adopt Area Under ROC Curve (AUC) to evaluate the AWI performance. 
AUC \cite{BRADLEY19971145} measures the discrimination degree of an AWI classifier. The value of AUC is ranged from 0 to 1. The random prediction has an AUC of 0.5 and a higher AUC indicates a better discrimination degree. 
In our study, the main reason for selecting AUC is that AUC is insensitive to the class imbalance \cite{ge2021impact}. 
Besides, AUC is also adopted alone for the performance evaluation in the previous ML-based AWI studies \cite{golden, a8, a10}.

% To describe AUC, we first define four basic terms. True Positive (TP) and False Negative (FN) describe the number of actionable warnings that are correctly and incorrectly identified by the AWI classifier respectively. True Negative (TN) and False Positive (FP) describe the number of unactionable warnings that are correctly and incorrectly identified by the AWI classifier respectively.
% AUC is denoted by the area under the ROC curve, which is created by plotting False Positive Rate (FPR = FP/(FP+TN)) and True Positive Rate (TPR = TP/(TP+FN)).
% AUC measures the discrimination degree of the AWI classifier. The value of AUC is ranged from 0 to 1. The random prediction has an AUC of 0.5 and a higher AUC indicates a better discrimination degree. 

% Such a characteristic can facilitate objectively evaluate the AWI model performance in the imbalanced warning dataset of Table \ref{tab:dataset}.

\subsection{Selection of PTMs}
We select the studied PTMs for AWI based on the following criteria. 
On the one hand, PTM is publicly available because we  fine-tune the model. As such, we exclude PTMs without the released source code, e.g., Codex \cite{codex} and GPT-3 \cite{brown2020language}.
On the other hand, PTM is trained on large-scale programming language corpora because AWI is a code-related task. As such, we exclude PTMs by only pre-training natural language texts, e.g., T5 \cite{raffel2020exploring} and GPT-2 \cite{radford2019language}.
At last, we select five representative PTMs (i.e., CodeBERT, GraphCodeBERT, CodeT5, UniXcoder, and CodeGPT). 
First, the five PTMs are widely used for various code-related tasks \cite{niu2023empirical}. 
Second, the five PTMs span different architectures (i.e., encoder-only, decoder-only, and encoder-decoder models) and organizations (i.e., Microsoft and Salesforce).
Third, the five PTMs are publicly accessible from Hugging Face \cite{huggingface}, which is by far the largest open-source large language model community.

\textbf{CodeBERT.} CodeBERT \cite{codebert} is a bimodal PTM to capture the semantic connection between Natural Language (NL) and Programming Language (PL) via the multi-layer and encoder-only transformer architecture. 
CodeBERT can learn general-purpose representations to support downstream NL-PL applications, e.g., the natural language code search.

\textbf{GraphCodeBERT.} GraphCodeBERT \cite{guo2020graphcodebert} a structure-aware PTM to track the inherent structure of source code based on the encoder-only transformer architecture. 
Unlike the existing PTMs that regard a code snippet as a sequence of tokens, GraphCodeBERT seizes crucial code semantics to enhance the code understanding process. 

% an extension of T5 architecture \cite{raffel2020exploring},
\textbf{CodeT5.} CodeT5 \cite{wang2021codet5} is obtained by utilize the encoder-decoder transformer to fine-tune T5 \cite{raffel2020exploring} for code-related tasks. 
Compared to T5, CodeT5 proposes an identifier-aware pre-training mechanism to convey code semantics from developer-assigned identifiers, which can help CodeT5 seamlessly support code understanding and generation tasks.

\textbf{UniXcoder.} UniXcoder \cite{guo2022unixcoder} is a unified cross-modal PTM for PL via the encoder-only transformer.
UniXcoder employs mask attention matrices with prefix adapters to dominate the model behavior and exploits cross-modal contents (e.g., AST and code comment) to enhance the code representation.

\textbf{CodeGPT.} CodeGPT \cite{lu2021codexglue} is a GPT-style PTM to tackle sequence-to-sequence generation tasks via an decoder-only transformer. Like GPT-2 \cite{radford2019language}, CodeGPT aims to predict the next token given all previous tokens.

% In particular, such a technique involves three different matching strategies (i.e., location-, snippet-, hash-based matching), which are placed for the  in order. 

\subsection{Dataset} \label{dataset}
% \dao{Change the subsection title to Dataset of Bug Warnings?}
We collect 10140 distinct SpotBugs warnings (i.e., 9666 unactionable and 474 actionable warnings) from 10 open-source and large-scale Java projects. TABLE \ref{tab:dataset} shows the dataset details. 
It is noted that SpotBugs involves ten warning categories, where each warning category contains multiple warning types.
The warning labeling process for each project is shown as follows. 
Given a project with a set of commits, we first filter out all compilable commits via \emph{Apache Maven} because SpotBugs can only run compilable commits. 
Then, we obtain a set of warnings by utilizing SpotBugs to scan the source code of each compilable commit. 
After that, we track the warning evolution among all compilable commits via a SOTA multi-stage warning matching technique \cite{mining}. 
The core idea behind such a technique is to conduct a pair-wise warning comparison between the pre-commit and post-commit by placing the location-, snippet-, and hash-based matching strategies in order. 
Once all warnings are compared among all compilable commits, such a technique initially labels these warnings to be closed, open, and unknown. 
To ensure the automatic warning label reliability, we follow the manual inspection criteria \cite{datasetheihei} to further confirm closed warnings into actionable, unactionable, and unknown ones. 
After that, the manually confirmed actionable and unactionable warnings are retained while the others are excluded. 
The number of open warnings is far more than that of closed warnings. Thus, inspired by the verification latency in software defects \cite{sdp-delay}, we calculate the lifetime of actionable warnings to further filter open warnings into unactionable and unknown ones. 
In particular, the lifetime of each unactionable/actionable warning is the time interval between the first occurrence of this warning and the persistence/disappearance of this warning \cite{observe}.
If the lifetime of an open warning is more than the median lifetime of actionable warnings, this warning is assigned to be unactionable and is retained. Otherwise, this warning is labeled to be unknown and is excluded. 
Finally, we can obtain the ground-truth actionable and reliable unactionable warnings.

% which are gather

% gather actionable and unactionable warnings as the final warning dataset 

% obtain sufficiently reliable labels for actionable and unactionable warnings.

% After that, the automatically filtered unactionable warnings are considered as to be sufficiently reliable.
% After that, the manually confirmed actionable and unactionable warnings are considered as to be ground-truth.
%
% including the manually confirmed actionable and unactionable warnings as well as the automatically filtered unactionable warnings.
% Only the manually confirmed actionable and unactionable warnings are retained, while the others are excluded.
% In our study, we focus on unactionable and actionable warnings for the evaluation. 
% The more details of such a technique can be seen in Liu et al. \cite{mining}. 

\subsection{Experimental Setup} \label{setup} 
In the SOTA ML-based AWI approach, we use PyTorch \cite{pytorch} to implement CNN and LSTM for AWI. The architecture design of CNN and LSTM follows the work of Lee. et al. \cite{sanxing} and Koc et al. \cite{empirical} respectively. 
For CNN, we set the word embedding dimension to 128, the batch size to 20, the dropout rate to 0.5, and use the SGD optimizer \cite{sgd} with 0.005 learning rate. For LSTM, we set the word embedding dimension to 8 and the batch size to 64.
In the PTM-based AWI approach, we use the Hugging Face \cite{huggingface} implementation version. Particularly, we set the batch size to 4, set the length of the input sequence to 256, and use the Adam optimizer \cite{kingma2014adam} with 5\emph{e} - 5 learning rate.
There could be multiple model architectures with different sizes in some PTMs (e.g., CodeT5-base and CodeT5-large). 
Since the base version is more practical and is employed with comparable effectiveness compared to the large version \cite{zhang2023pre}, we select PTMs with the base version for AWI.

As for RQ1$\sim$3 and RQ5, we merge all warnings from 10 projects and conduct the stratified sampling based on the ratio of 7/1/2. That is, 70\%, 10\%, and 20\% of all warnings are split into training, validation, and testing sets respectively. 
As for RQ4, we conduct the stratified sampling for warnings in each project based on the ratio of 1/1. That is, 50\% of warnings are taken as the training set and the remaining 50\% of warnings are taken as the test set. 
Due to no validation set to support the final parameter determination in RQ4, we set the epoch to 30 in the model training of PTMs.
Particularly, due to the class imbalance in all warnings, we adopt stratified sampling rather than random sampling, so as to ensure that each respective set contains actionable warnings.
In addition, all experiments are conducted with one Ubuntu 18.04.3 server with two Tesla V100-SXM2 GPUs. 
% As for the studied PTMs, we use the Hugging Face \cite{huggingface} implementation version with default parameters. 
% There could be multiple model architectures with different sizes in some PTMs (e.g., CodeT5-based and CodeT5-large). 
% Since the base version is more practical and is employed with comparable effectiveness than the large version \cite{zhang2023pre}, we select the base version for the evaluation. 

% Besides, all experiments adopt AUC to evaluate the AWI performance.
% Intel(R) Core(TM) i7-10810U CPU @ 1.10GHz   1.61 GHz 32GB
\section{Results and Analysis} \label{result} 

\begin{figure}
    \centering
    \includegraphics[scale=0.6]{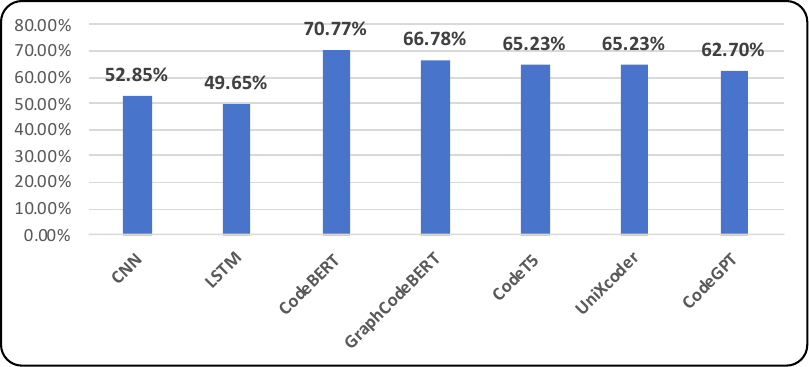}
    \caption{AUC of DL-based and PTM-based AWI approaches.}
    \label{fig:rq1}
\end{figure}
% \subsection{RQ1: How is the performance of PTMs on AWI in comparison to the state-of-the-art ML-based AWI approach?} 
\subsection{RQ1: Effectiveness of PTM-based AWI}
\label{rq1}

\textbf{Motivation.} This RQ aims to explore the effectiveness of PTM-based AWI. Besides, this RQ investigates the performance differences of different PTMs on AWI.

% 解释为什么选用作为state-of-the-art baseline
% Based on the classification of adopted ML models \cite{ge2023machine}, the existing ML-based AWI approaches can be divided into the traditional ML-based and DL-based AWI approaches. 
% Due to the powerful ability of DL techniques in the feature representation, DL-based AWI approaches generally achieve higher performance than the traditional ML-based AWI approaches \cite{empirical, empiricalex}. 
% In general, DL-based AWI approach takes the warning context as the natural language text and uses DL models to learn an AWI classifier from the warning context.
\textbf{Design.} To answer this RQ, we compare the SOTA ML-based with PTM-based AWI approaches.
As described in Section \ref{background}, the DL-based AWI approach performs better than the traditional ML-based AWI approach in the existing ML-based AWI approaches \cite{empirical, empiricalex, S44}.
Thus, we select the DL-based AWI approach as the SOTA ML-based AWI approach.
% 具体的细节
Following previous DL-based AWI studies \cite{empirical, empiricalex, smartcontract, MAPL, sanxing}, we extract source code from the method containing a warning as the warning context, abstract identifiers and literals in the warning context, and use the DL model to train a classifier for AWI on the abstracted warning context. 
Instead of directly using the source code related to warning line numbers for AWI, we extract the warning context from the method containing a warning for AWI. It is because previous studies \cite{MAPL, sanxing, smartcontract} signify that compared to the warning line numbers, the method containing a warning can provide more traceability information to judge whether a warning is actionable or not. 
Besides, we abstract each identifier/literal in the warning context with a unique ID, and the detailed warning context abstraction process can be seen in Section \ref{rq22}.
Moreover, we attempt CNN and LSTM for AWI because the results of previous studies \cite{empirical, empiricalex, sanxing, smartcontract} demonstrate the superior performance of CNN and LSTM on AWI. 
% PTM-based AWI  
In the PTM-based AWI approach, we select the optimal AUC for comparison.
As shown in Section \ref{rq1} and \ref{rq2}, when extracting the warning context from the method containing a warning, having no warning context abstraction, and using the pre-training and fine-tuning components, the PTM-based AWI approach can obtain the optimal performance. 
\textbf{Results.} Fig. \ref{fig:rq1} shows that in terms of AUC, CodeBERT, GraphCodeBERT, CodeT5, UniXcoder, and CodeGPT outperform CNN by 17.92\%, 13.93\%, 12.38\%, 12.38\%, and 9.85\% as well as LSTM by 21.12\%, 17.13\%, 15.58\%, 15.58\%, and 13.05\% respectively. 
It signifies that PTM can substantially improve the SOTA AWI performance.
By further investigation, there are two possible factors that the AUC improvement of PTMs over DL models on AWI. 
On the one hand, PTMs leverage extensive codebases to yield more significant vector representation. For example, CodeBERT has 2.1M bimodal and 6.4M unimodal code-related datapoints across six programming languages. 
By contrast, CNN and LSTM are trained on a limited dataset with 70\% (7089) of all warnings. 
On the other hand, PTMs employ the transformer architecture, which can provide the context for any position in a given input sequence via the self-attention mechanism. 
However, CNN and LSTM cannot capture the relative position information due to the absence of the transformer architecture.

% Thus,  CodeGPT underperforms CodeBERT, GraphCodeBERT, and UniXcoder in the AWI. 
% the separate code classification task. 
Fig. \ref{fig:rq1} presents that in terms of AUC, the encoder-only PTMs (i.e., CodeBERT, GraphCodeBERT, UniXcoder) are basically better than the remaining two PTMs (i.e., the encoder-decoder CodeT5 and the decoder-only CodeGPT). 
The main reasons for this phenomenon are shown as follows. 
AWI aims to perform the warning classification by understanding the warning context. 
Exactly, the encoder-only PTMs are very good at handling the code classification tasks due to considering the contextual information of source code.
In contrast, despite being able to handle the code classification and generation tasks, CodeT5 juggles both encoder and decoder parts, which could cause suboptimal performance in AWI.
The decoder-only PTMs, focusing on code generation tasks, could exhibit a restricted ability in AWI due to not fully leveraging the contextual information of warnings \cite{llmclassification}. 
Besides, CodeBERT obtains the optimal AUC of 70.77\% in the three encode-only PTMs. It indicates that CodeBERT is 3.99\% and 5.54\% better than GraphCodeBERT and UniXcoder respectively. 
We speculate that such a slight performance difference may be caused by corpora with different scales and code datapoints.

\begin{tcolorbox}[colback=gray!13, colframe=black, boxrule=0.3mm, boxsep= -0.1cm, middle=-0.1cm]
\textbf{Answering RQ1}: PTMs exhibit remarkable AWI performance, which substantially outperforms the state-of-the-art ML-based AWI approach by 9.85\%$\sim$21.12\% in terms of AUC. Besides, CodeBERT achieves the optimal AUC of 70.77\% among the five studied PTMs.
\end{tcolorbox}

% \subsection{RQ2: How do the data preprocessing ways affect the performance of PTMs on AWI?} 
\subsection{RQ2: Analysis in the data preprocessing} 
\label{rq2}
% This section investigates how different data preprocessing ways affect the performance of PTMs on AWI. 
% Based on the typical PTM-based AWI workflow in Section \ref{overview}, this RQ explores the impact of the warning context construction and abstraction on the PTM-based AWI approach.

% In RQ1, we only feed the source code, extracted from the method containing a warning, into the DL-based and PTM-based AWI approaches. 
% \subsubsection{RQ2.1} What is the impact of warning context extraction?

\textbf{Motivation.} 
Based on the typical PTM-based AWI workflow in Section \ref{overview}, the data preprocessing contains the warning context extraction and abstraction.
The warning context is the source code related to the warning line numbers. 
Previous studies \cite{MAPL, sanxing, smartcontract} show that the warning context plays a crucial role in AWI. 
The warning context abstraction is to rename raw code tokens to a set of predefined tokens, thereby reducing the number of code tokens in the warning context.
The existing DL-based AWI studies explicitly demonstrate that the abstracted warning context is beneficial for AWI \cite{empirical, empiricalex} compared to the raw warning context.
However, the impact of both warning context extraction and abstraction on the PTM-based AWI approach has not been fully investigated yet. 
Thus, this RQ explores the performance of the PTM-based AWI approach in different data preprocessing ways.

\subsubsection{RQ2.1} The impact of warning context extraction. \label{rq21}

\textbf{Design.} 
% To answer this RQ, we perform PTM-based AWI on with or without the warning context. 
Given a warning reported by SpotBugs, the warning location can be obtained, including the class/method information containing this warning as well as the warning line numbers.
As for each warning, we extract source code via the warning line numbers, which is called without the warning context.
By contrast, we extract source code from the method containing a warning, which is called with the warning context. 
Not all warnings, especially for warnings related to class member variables, are reported inside methods. 
As such, the context of a warning outside the method is extracted via the warning line number. 
% It is noted that the warnings outside methods only account for a very small ratio of all warnings. 
%
It is noted that instead of extracting the warning context from the class containing a warning, we extract the warning context from the method containing a warning.
There are three main reasons. 
First, the existing studies \cite{vrepair, gra2} have observed that defects are generally revealed by analyzing source code in the method scope.
Second, a previous study \cite{MAPL} demonstrates that it is proper to extract the warning context for AWI in the method granularity. 
Third, the class granularity could bring too much noise into the warning context compared to the method granularity.

% In all, this RQ involves 10 experiments (five PTMs * two extraction settings). 
% In each experiment, we follow the experiment settings of RQ1. That is, after the stratified sampling, we use 70\%, 10\%, and 20\% of all warnings as the training, validation, and test sets respectively. 
% To the end, we adopt the AUC to evaluate AWI performance difference between the two settings.

% using source code extracted from the method containing a warnings over the source code extracted from the warning line numbers in the .
% % On average, the pTM-based AWI approach with the warning context is 3.37\% higher than that without the warning context in terms of AUC.
% It indicates that the pTM-based AWI approach with the warning context consistently outperforms that without the warning context. 
% In particular,  CodeBERT on AWI reaches a new record of 70.77\%, exceeding CNN and LSTM on AWI by 17.92\% and 21.12\% respectively. 

\begin{figure}
    \centering
    \subfloat[The warning context extraction.]{
    \label{fig:rq21}
    \includegraphics[scale=0.6]{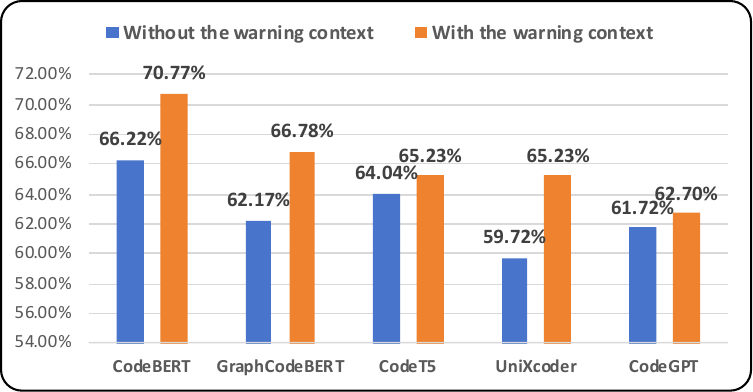}}
    \hfill
    \subfloat[The warning context abstraction.]{
    \label{fig:rq22}
    \includegraphics[scale=0.6]{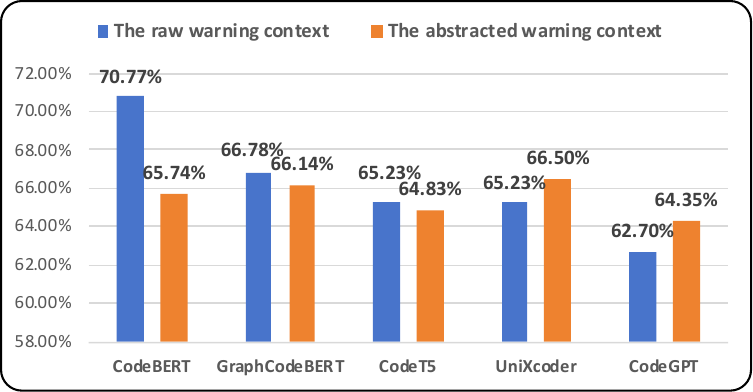}}
    \caption{AUC of PTMs on AWI in the data preprocessing.}
    \label{fig:rq2}
\end{figure}

% The above results highlight the substantial benefits of the warning context for the PTM-based AWI approach. 
% In particular,  CodeBERT on AWI reaches a new record of 70.77\%, exceeding CNN and LSTM on AWI by 17.92\% and 21.12\% respectively. 
\textbf{Results.} As presented in Fig. \ref{fig:rq21}, CodeBERT, GraphCodeBERT, CodeT5, UniXcoder, and CodeGPT with the warning context are 4.55\%, 4.61\%, 1.19\%, 5.51\%, 0.98\% better than that without the warning context respectively. 
On average, the pTM-based AWI approach with the warning context is 3.37\% higher than that without the warning context in terms of AUC.
It highlights substantial benefits of the warning context for the PTM-based AWI approach.  
Further investigation shows the possible reason for such a phenomenon.
Without the warning context, the source code extracted from warning line numbers could only denote the appearance of warnings. 
With the warning context, in addition to involving the source code extracted from the warning line numbers, the source code extracted from the method containing a warning could embrace the root cause of a warning, which greatly bolsters the PTM-based AWI approach.

% \subsubsection{RQ2.2} What is the impact of warning context abstraction?
\subsubsection{RQ2.2} The impact of warning context abstraction. \label{rq22}

% \textbf{Motivation.} The warning context abstraction is to rename raw code tokens to a set of predefined tokens, thereby reducing the number of code tokens in the warning context.
% The existing DL-based AWI studies explicitly demonstrate that the abstracted warning context is beneficial for AWI \cite{empirical, empiricalex} compared to the raw warning context. However, the impact of warning context abstraction on the PTM-based AWI approach has not been explored yet. 
% Thus, this RQ analyzes whether the PTM-based AWI approach performs differently with or without the warning context abstraction.

% 这儿直接从抽象说起，是否会存在的歧义
% 为什么只是warning line number， warning method，而没有warning所在file
% Then, we design four different regular expressions (i.e., ``.*?\*/'', ``/\*.*", ``/\*.*?\*/", and ``//.*") to remove comments in the warning context.
\textbf{Design.} To answer this RQ, we compare the performance difference of the PTM-based AWI approach between the raw and abstracted warning contexts.
As for the raw warning context, we directly use source code extracted from the method containing a warning. 
As for the abstracted warning context, we first extract source code of the method containing a warning as the warning context. 
Then, we tokenize the warning context via the Java lexical analysis.
After that, by utilizing JavaParser \cite{javaparse} to construct an Abstract Syntax Tree (AST), we identify identifier and literal types from the warning context. 
Finally, we replace each identifier/literal in the stream of tokens with a distinct number, which denotes the type and role of this identifier/literal in the warning context.
For example, the source code ``int a = 1;'' is abstracted into ``int intVar1 = intLiteral1;''. 
% Finally, by matching AST node types (i.e., MethodDeclarition and FileDeclarition) between the raw source code extracted from the methods containing warnings and the abstracted source code extracted from files containing these warnings, we obtain the abstracted warning context from the methods containing warnings.

% In all, this RQ involves 10 experiments (five PTMs * two abstraction settings). 
% Following the experiment settings of RQ1, we use 70\%, 10\%, and 20\% of all warnings from the stratified sampling as the training, validation, and test sets in each experiment respectively. 

% CodeBERT, GraphCodeBERT, and CodeT5 with the raw warning context are better than those with the abstracted warning context. Conversely, UniXcoder and CodeGPT with the raw warning context are inferior to those with the abstracted warning context.
\textbf{Results.} Fig. \ref{fig:rq22} shows that in terms of AUC, the raw warning context improves the abstracted warning context by 5.03\% in CodeBERT, 0.64\% in GraphCodeBERT, 0.40\% in CodeT5 respectively. 
However, UniXcoder and CodeGPT with the abstracted warning context outperform those with raw warning context by 1.27\% and 1.65\% respectively. 
The AWI performance differences among five PTMs could be caused by the number of training corpora, different modalities, and architectures of PTMs \cite{niu2023empirical}. 
On average, the AUC of the raw warning context is 0.63\% higher than that of the abstracted warning context across five PTMs.
Such an observation is in violation of that in the existing DL-based AWI studies \cite{empirical, empiricalex}. 
By further analysis, the possible reason is shown as follows. 
In the DL-based AWI approach, the abstracted warning context can significantly reduce the size of code tokens, thereby facilitating improving the AWI generalizability.
However, in the PTM-based AWI approach, the abstracted warning context may hide valuable information about the raw works that can be learned by word embedding. 
Thus, the AWI performance could decrease when the abstracted warning context serves as the input of certain PTMs.

% Overall, the raw warning context is more beneficial for the PTM-based AWI approach than the abstracted warning context.

% By further analysis, the possible reason could be different modalities of the PTMs. 
% CodeBERT, GraphCodeBERT, and CodeT5 are obtained by jointly pre-training on the code and document.
% By contrast, UniXcoder and CodeGPT are obtained by pre-training either the code or document. 
% The abstracted warning context may be insensitive to when jointly pre-training on the code and document.
% Thus, such the observation spurs researchers and practitioners to further delve the impact of warning context construction in a specific PTM for AWI.

\begin{tcolorbox}[colback=gray!13, colframe=black, boxrule=0.3mm, boxsep= -0.1cm, middle=-0.1cm]
\textbf{Answering RQ2}: The performance of the PTM-based AWI approach is affected by data preprocessing ways.
In detail, the PTM-based AWI approach with the warning context consistently outperforms that without the warning context. 
The warning context abstraction may hinder the utilization of PTM's generic knowledge on AWI, which causes a slight decline of AWI performance in certain PTMs compared to the raw warning context.
\end{tcolorbox}

% \subsection{RQ3: How do the model training components affect the performance of PTMs on AWI?} 
\subsection{RQ3: Analysis in the model training} 
\label{rq3} 
% This section investigates how different model training components affect the performance of PTMs on AWI. Based on the typical PTM-based AWI workflow in Section \ref{overview}, this RQ explores the contribution of pre-training and fine-tuning components to the PTM-based AWI approach. 

\textbf{Motivation.} The results in RQ1 show that the PTM-based AWI approach outperforms the SOTA ML-based AWI approach. The ML-based AWI approach is trained in a traditional pipeline, i.e., supervised learning on labeled warnings. 
In contrast, the PTM-based AWI approach involves two components in the model training, including a pre-training component for a general task with self-supervised learning on large-scale corpora and a fine-tuning component for a downstream task with supervised learning on labeled warnings. 
Thus, this RQ investigates the role of the pre-training and fine-tuning components when using PTMs for AWI. 
% It is necessary to explore the role of two components in the PTM-based AWI approach. 
% This RQ separately investigates the role of the pre-training component when using PTMs for AWI. 

\subsubsection{RQ3.1} The role of a pre-training component. 
\label{rq31}

\textbf{Design.} Based on the classification of PTM architectures in Section \ref{ptm}, we select three models (i.e., the encoder-only BERT, the encoder-decoder T5, and the decoder-only GPT) as baselines without a pre-training component. 
Correspondingly, we select CodeBERT/CodeT5/CodeGPT, which are obtained by pre-training BERT/T5/GPT on massive codebases respectively, as PTMs with a pre-training component. 
Since the results in Section \ref{rq21} show that PTMs with the raw warning context achieve the optimal performance on AWI, we select the raw warning context for evaluation in this RQ.
% In all, there are six experiments (i.e., two pre-training variants * three PTMs). 
% Following the experiment settings of RQ1, we use 70\%, 10\%, and 20\% of all warnings from the stratified sampling as the training, validation, and test sets in each experiment respectively. 

\begin{figure}
    \centering
    \includegraphics[scale=0.6]{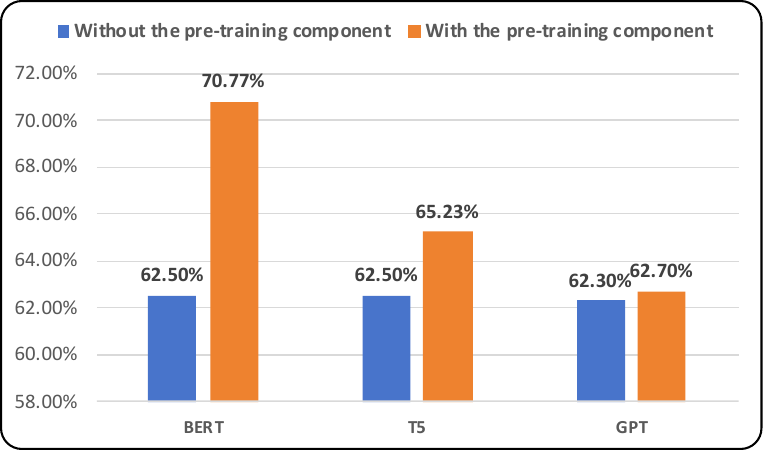}
    \caption{AUC of PTMs on AWI with/without pre-training.}
    \label{fig:rq31}
\end{figure}

%  gpt 65.48
\textbf{Results.} As shown in Fig. \ref{fig:rq31}, CodeBERT, CodeT5, and CodeGPT achieve the AUC of 70.77\%, 65.23\%, and 62.70\% respectively. 
By contrast, BERT, T5, and GPT only obtain the AUC of 62.50\%, 62.50\%, and 62.30\% respectively. 
Regardless of PTMs, the pre-training component consistently improves the AWI performance by 0.40\%$\sim$8.27\%. 
It signifies that the pre-training component can provide substantial benefits for the PTM-based AWI approach. 
Besides, it is observed in Fig. \ref{fig:rq31} that for PTMs without a pre-training component, BERT outperforms GPT by 0.20\%.
Such an observation underlines that the encoder-only model could be more suitable for code classification tasks (i.e., AWI) than the decoder-only model.

% the encode-only model can generally achieve the highest AWI performance, followed by the encoder-decoder and decoder-only models. 
% is obviously reflected in the PTMs with a pre-training component. 

% That is, CodeBERT outperforms CodeT5 and CodeGPT by 5.54\% and 8.07\% respectively. Also, CodeT5 exceeds CodeGPT by 2.53\%. 

\begin{figure*}
    \centering
    \includegraphics[scale=0.52]{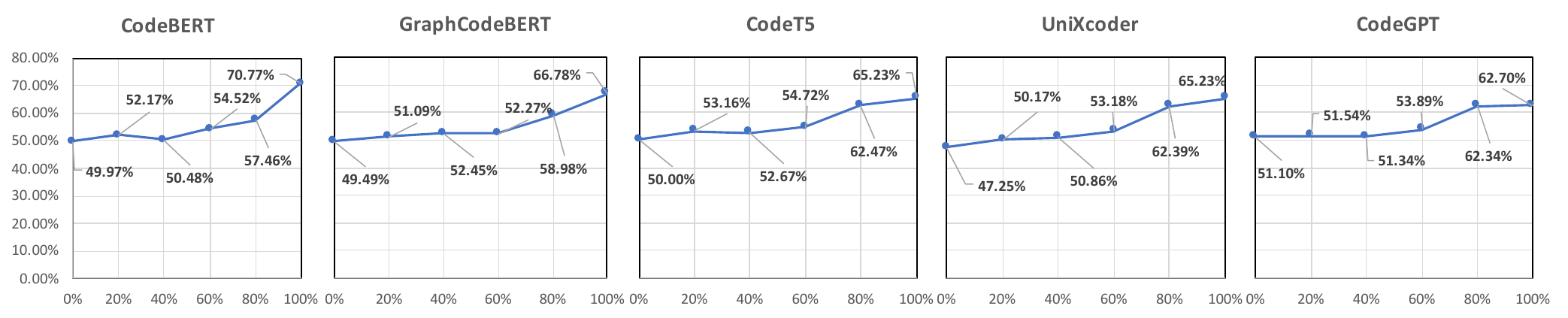}
    \caption{AUC trend of PTMs on AWI with different fine-tuning corpora.}
    \label{fig:rq32}
\end{figure*}

% \subsubsection{RQ3.2: What is the role of a fine-tuning component?} 
% \subsubsection{RQ3.2} What is the role of a fine-tuning component? 
\subsubsection{RQ3.2} The role of a fine-tuning component. 
\label{rq32}

% \textbf{Motivation.} As shown in the motivation of Section \ref{rq31}, this RQ separately investigates the role of a fine-tuning component in the PTM-based AWI approach. 

\textbf{Design.} In RQ1, we rely on the stratified sampling to select 70\%, 10\%, and 20\% of all warnings as the training, validation, and test sets respectively. 
To answer this RQ, we select 0$\sim$100\% warnings from the training set with 20\% intervals each time. It indicates that there are six fine-tuning corpora (i.e., 0\%, 20\%, 40\%, 60\%, 80\%, and 100\% warnings of the training set). 
After that, we train the PTM-based AWI classifier on each fine-tuning corpus, determine the optimal parameters of this classifier on the validation set, and evaluate this classifier on the test set. 
Similar to Section \ref{rq31}, we select the raw warning context for evaluation in this RQ.
% In all, there are 30 experiments (six fine-tuning corpora * five PTMs). 
% Noted, due to the optimal performance of the raw warning context in Section \ref{rq2}, we select the raw warning context for evaluation in this RQ.

\textbf{Results.} Fig. \ref{fig:rq32} shows that as the number of fine-tuning corpora increases, the AUC of the PTM-based AWI approach has an upward trend. 
When the percentage of fine-tuning corpora in the training set increases from 0\%$\sim$60\%, the AUC of five PTMs on AWI is slowly rising. 
Also, such a rising trend of AUC could be unstable in some PTMs. For example, when the fine-tuning corpora account for 20\% of the training set, the AUC of CodeBERT on AWI is 52.17\%. 
However, when the fine-tuning corpora increase to 40\% of the training set, the AUC of CodeBERT on AWI has a slight decline with 1.69\%. 
By further analysis, the possible reason is that due to randomly selecting fine-tuning corpora from the imbalanced training set in the early stage, the corpora fed to PTMs could be almost unactionable warnings. If there are few actionable and massive unactionable warnings in the fine-tuning corpora, PTMs may occasionally underperform on AWI.
When the fine-tuning corpora reach up to 60\% of the training set, taking the results of Fig. \ref{fig:rq1} into account, the AUC of PTMs is comparable or even higher than that of CNN/LSTM on AWI. 
It indicates that the number of fine-tuning corpora plays a crucial role when using PTMs for AWI.

Besides, five PTMs with no fine-tuning corpora show poor AWI performance.
Despite acquiring valuable knowledge from the pre-training component, these PTMs could not adapt to the downstream task (i.e., AWI) without a fine-tuning process. 
Particularly, when the fine-tuning corpora are 100\% of the training set, the AUC of PTMs on AWI still has no trend to slow down. 
Such findings further underscore the advantages of the fine-tuning component in the PTM-based AWI approach, which can enable PTMs to acquire task-specific expertise and maximize the utilization of the knowledge gained from the pre-training component.

% In further research, it is very essential to explore the performance of the PTM-based AWI approach on a larger and richer warning dataset.
% Future research can be conducted to explore the direct usage of PTMs for AWI in a zero-shot setting. For example, we can replace the vulnerable line with a mask line and query the models to fill the mask line with replacement tokens to produce candidate patches

\begin{tcolorbox}[colback=gray!13, colframe=black, boxrule=0.3mm, boxsep= -0.1cm, middle=-0.1cm]
\textbf{Answering RQ3}: The performance of the PTM-based AWI approach is affected by the model training components. In detail, the pre-training component can acquire general knowledge from codebases to further enhance AWI. 
The performance of PTMs on AWI is gradually rising with the increase of fine-tuning corpora. 
It indicates that the pre-training and fine-tuning components play a key role in the PTM-based AWI approach.
\end{tcolorbox}

% \subsection{RQ4: How do the model prediction scenarios affect the performance of PTMs on AWI?} 
\subsection{RQ4: Analysis in the model prediction} 
\label{rq4}

\textbf{Motivation.} The typical PTM-based AWI workflow in Section \ref{overview} shows that the model prediction involves two scenarios, i.e., within and cross project AWI.
As shown in Section \ref{rq1}, the results on all warnings of 10 projects describe that the PTM-based AWI approach performs better than the SOTA ML-based AWI approach.
However, little work explores how PTMs perform in within and cross project AWI scenarios, which fails to understand the performance difference of PTMs in different model prediction scenarios.
Thus, this RQ aims to bridge the above gap.
% aims to investigate how two model prediction scenarios affect the performance of PTMs on AWI.  

% little work attempts to investigate the performance differences of PTMs between within and cross project AWI scenarios. 

% In the AWI field, there are two typical evaluation scenarios, i.e., the within project and cross project AWI. 
% In the pre-trained context, the within-project AWI refers that the training set (i.e., the fine-tuning corpora) and the test set are from the warning dataset with the same project. 
% By contrast, the cross-project AWI refers that the training set (i.e., the fine-tuning corpora) and the test set are from the warning dataset with different projects.
% The results in Section \ref{rq1} demonstrate that the PTM-based AWI approach obviously outperforms the DL-based AWI approach in the warning dataset.
% However, it is unclear that how the PTM-based AWI approach performs in the within-project and cross-project AWI. Also, it is unknown what differences the PTM-based AWI approach is between the within-project and cross-project AWI. 
% Thus, this RQ investigates the performance of the PTM-based AWI approach in the within-project and cross-project AWI. 

% we use the stratified sampling to equally divide warnings in each project. That is, 50\% of warnings are taken as the training set and the remaining 50\% of warnings are taken as the test set. 

\textbf{Design.} TABLE \ref{tab:dataset} shows that warnings are collected from 10 projects. 
To answer this RQ, we first use the stratified sampling to take 50\% of warnings as the training set and take the remaining 50\% of warnings as the test set in each project. 
After that, we construct within and cross project AWI scenarios, which are shown in TABLE \ref{tab:rq4}. 
There are two variants in the within and cross project AWI scenarios respectively. The difference between within1 (cross1) and within2 (cross2) is whether the training set contains warnings from the remaining nine projects when taking 50\% of warnings in a project as the test set. 
Besides, to conduct a fair comparison between within and cross project AWI scenarios, the test set is the same in four variants.
Such a rigorous design aims to further investigate whether the number of training set affects the performance of PTMs in within and cross project AWI scenarios. 
Similar to Section \ref{rq31}, we select the raw warning context for evaluation in this RQ.
% In all, there are 200 experiments (four variants * 10 projects * five PTMs). 
% Since the results in RQ2.1 show that PTMs with the raw warning context achieve the optimal performance on AWI, we select the raw warning context for evaluation in this RQ.

\begin{table}[]
    \centering
    \begin{tabular}{|p{20pt}<{\centering}|p{20pt}<{\centering}|p{120pt}<{\centering}|p{40pt}<{\centering}|}
    \hline
         \multicolumn{2}{|c|}{\textbf{AWI scenario}}		&		\textbf{Training set}	&	\textbf{Test set}	\\	\hline \hline
\multirow{2}{20pt}{Within project}	&	Within1	&	All warnings in nine projects and 50\% of warnings in the tenth project	&	\multirow{4}{40pt}{Remaining 50\% of warnings in the tenth project}		\\ \cline{2-3}
% \rowcolor{gray!40}
	&	Within2	&	50\% of warnings in the tenth project	&	\\	\cline{1-3}
\multirow{2}{20pt}{Cross project}	&	Cross1	&	All warnings in nine projects	&			\\ \cline{2-3}
	&	Cross2	&	N/A	&		\\	\hline
    \end{tabular}
    \caption{Within and cross project AWI scenarios.}
    \label{tab:rq4}
\end{table}

\begin{figure}
    \centering
    \includegraphics[scale=0.6]{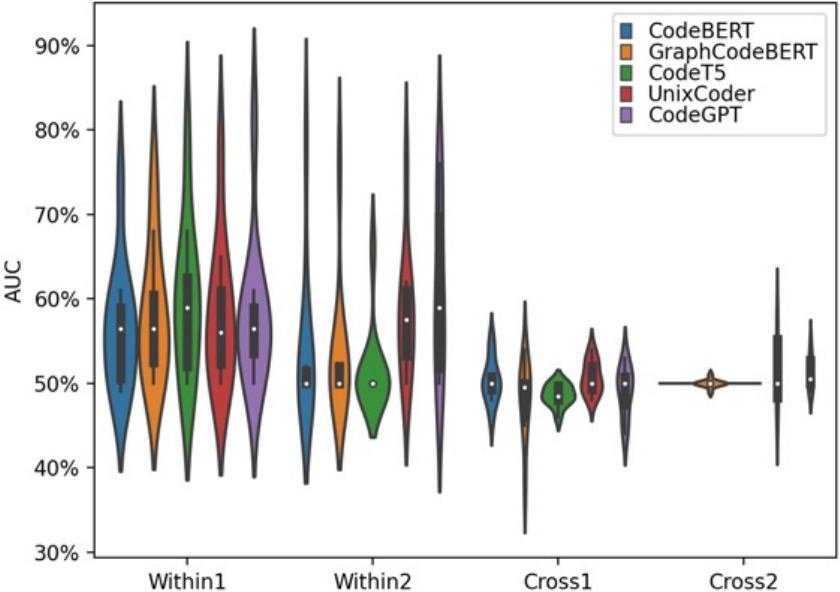}
    \caption{AUC of PTMs on AWI with different scenarios.}
    \label{fig:rq4}
\end{figure}

\textbf{Results.} Fig. \ref{fig:rq4} shows that PTMs in the within project AWI scenario obviously perform better than in the cross project AWI scenario. 
In five PTMs, the median AUC of Within1 is nearly 60.00\%, while the median AUC of Cross1 and Cross2 is only close to 50.00\%. 
It indicates that PTMs barely work in the cross project AWI scenario.
The main reason could be that the training and test sets are highly homogeneous in the within project AWI scenario due to coming from the same project.
Conversely, the training and test sets are heterogeneous in the cross project AWI scenario due to coming from different projects. 
Thus, the AWI-related expertise of PTMs gained by fine-tuning over the training set can work in within project AWI scenario but not in cross project AWI scenario.
Additionally, Within1 almost outperforms Within2 while Cross1 is similar to Cross2 in the median AUC of five PTMs. 
It signifies that the increased number of training set can greatly improve the PTM-based AWI approach in the within project AWI scenario. 
However, the improvement of the PTM-based AWI approach by increasing the number of training set is very limited because there is heterogeneity in the cross-project AWI scenario. 

\begin{tcolorbox}[colback=gray!13, colframe=black, boxrule=0.3mm, boxsep= -0.1cm, middle=-0.1cm]
\textbf{Answering RQ4}: The PTM-based AWI approach in the within project AWI scenario outperforms that in the cross project AWI scenario. 
Besides, the larger number of training sets can better enhance the PTM-based AWI approach in the within-project AWI scenario while only bringing limited improvement to the PTM-based AWI approach in the cross-project AWI scenario.
\end{tcolorbox}

\subsection{RQ5: Further analysis of incorrect predictions in the current PTM-based AWI } \label{rq5}
\textbf{Motivation.} Despite showing superior performance compared to the SOTA ML-based AWI approach in Sections \ref{rq1}$\sim$\ref{rq3}, the PTM-based AWI approach only achieves the optimal AUC of 70.77\%. It indicates that the PTM-based AWI approach still has substantial room for improvement. Thus, this RQ analyzes the underperformance of current PTMs on AWI, thereby providing future improvement directions.
% for future PTM-based AWI research

\begin{figure}
    \centering
    \includegraphics[scale=0.5]{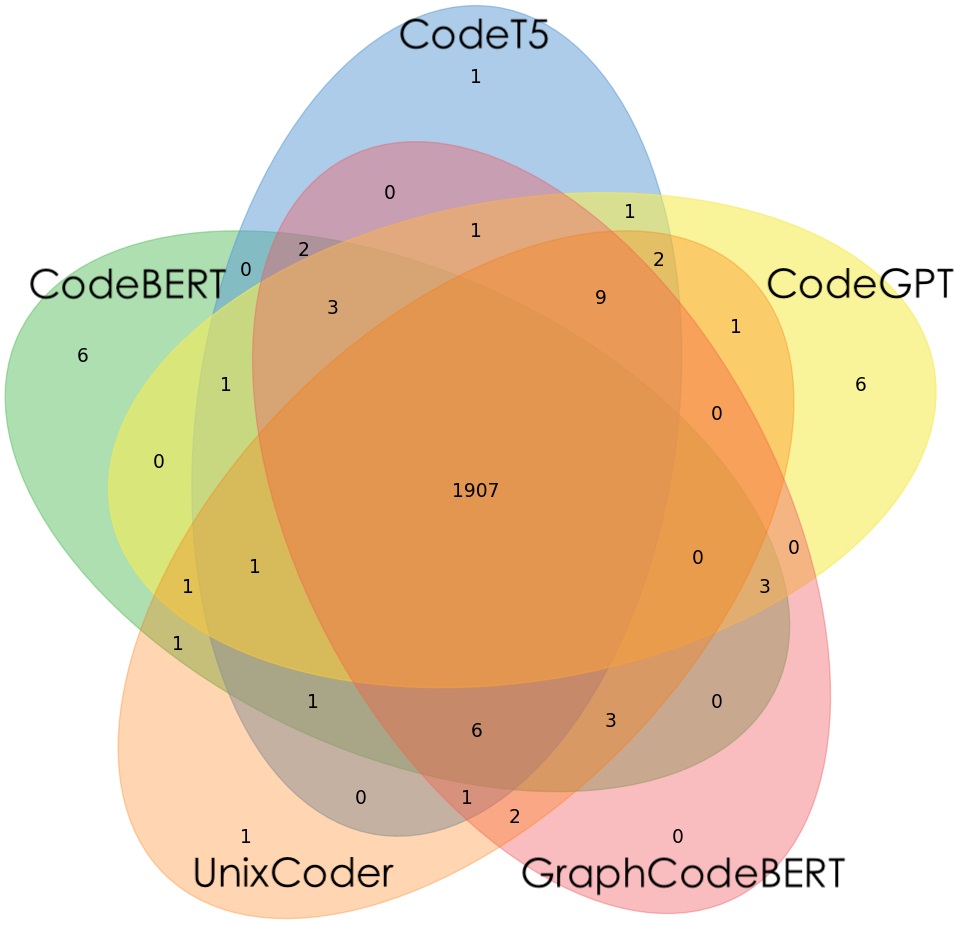}
    \caption{Venn diagrams of the number of correctly classified warnings five PTMs on the test set.}
    \label{fig:rq5}
\end{figure}

\textbf{Design.} Based on the results of PTMs on AWI in Section \ref{rq1}, we first gather the predicted labels of five PTMs in the test set (i.e., 2027 warnings). We then compare ground truth and predicted labels to confirm which warnings are wrongly classified by the PTM-based AWI approach. Finally, we analyze the reasons why PTMs fail to classify warnings.

% Fig. \ref{fig:rq5} illustrates the distribution of correctly classified warnings for five PTMs on the test set. In all, the final AUC that gathers five PTMs
\textbf{Results.} As shown in Fig. \ref{fig:rq5}, 1907 warnings are correctly classified by five PTMs. 
Also, a few warnings (e.g., 6 warnings) are only identified by some PTMs (e.g., CodeBERT). 
By gathering all correctly classified warnings from five PTMs, the final AUC is 72.94\%. It reflects that 68 warnings are wrongly classified by five PTMs. Further, we analyze and summarize reasons why PTMs cannot identify 68 warnings.

\textbf{Similar or even the same contexts in actionable and unactionable warnings.}
It is observed that warnings in the same type, where the warning type especially falls into categories with \emph{$BAD\_PRACTICE$}, \emph{$DODGY\_CODE$}, and \emph{$I18N$}, tend to have similar contexts. In these warnings, only a tiny part of warnings are acted on and fixed by developers. In contrast, most warnings are ignored by developers due to generally not affecting the functional correctness of the program. 
In addition, multiple warnings may appear in the same method, where some are actionable and others are actionable. However, based on the warning context extraction of our study, warnings in the same method have the same warning context. 
Thus, the above phenomena may cause PTMs to give ambiguous warning labels. 

% Such a phenomenon is especially obvious in the warning types, which fall into warning categories with \emph{$BAD\_PRACTICE$}, \emph{$DODGY\_CODE$}, and \emph{$I18N$}.
% Warnings with the same type usually have similar contexts, where some are actionable while others are unactionable. 
% Such a phenomenon is especially obvious in SpotBugs warning categories related to \emph{$BAD\_PRACTICE$}, \emph{$DODGY\_CODE$}, and \emph{$I18N$}. 
% Despite not being handled by developers, warnings in these categories generally do not affect the functional correctness of software. 
% In addition, there are multiple warnings in the same method, where some are actionable and others are actionable. However, warnings in the same method have the same warning contexts in the experimental design of our study. 
% Thus, PTMs could give ambiguous answers to these warnings in the above 

\textbf{Insufficient or unavailable warning contexts.} It is found that insufficient or unavailable warning contexts could cause PTMs to acquire inferior performance on AWI. 
Such a phenomenon is mainly reflected in the following aspects.
First, some warnings are from the methods with a single statement, e.g., a getter method that often returns a class field. 
The contexts of these warnings fail to offer insights into the potential information that a class field might contain.
Second, some warnings are related to the interprocedural method calls. Thus, it could not be enough to only extract the source code in the method containing a warning as the warning context in our study.
Third, some warnings fall into the methods with an interface type. The contexts of these warnings cannot be determined when the program is not executed.
Fourth, some warnings are related to the declaration or initialization of the class fields. As such fields are often outside a method, the contexts of these warnings are usually unavailable. 

% When the warning context is insufficient or unavailable, PTMs acquire inferior performance on AWI. 
% To ensure that there are actionable warnings in the training set, validation, and test sets respectively, our study adopts the stratified sampling to split the warning dataset. 
\textbf{Lacking adequate warnings with diverse types in the training set.} Our study fine-tunes PTMs on the training set and uses the fine-tuned PTMs for the test set. 
However, it is observed that some types of warnings (e.g., \emph{MS\_OOI\_PKGPROTECT} and \emph{RR\_NOT\_CHECKED}) in the test set are not included in the training set. 
Also, some types of warnings are extremely imbalanced in the training set. For example, all warnings with the \emph{MS\_MUTABLE\_ARRAY} type are actionable in the training set.
Thus, the above phenomenon in the training set may make PTMs learn a biased AWI classifier, which could not work well in the test set \cite{krawczyk2016learning}.

\section{Discussion} \label{discussion}

% \textbf{Further improvement for PTM-based AWI.} 
% Despite the performance of PTMs on AWI over the state-of-the-art ML-based AWI approach in Section \ref{rq1}, the optimal AUC of the PTM-based AWI approach is only about 70\%. 
% Besides, Section \ref{rq4} demonstrates that PTMs are underperforming on the within and especially cross project AWI scenarios. That is, the median AUC is nearly 60\% in the within project AWI scenario while only being close to 50\% in the cross project AWI scenario. 
% Moreover, Sections \ref{rq2} and \ref{rq3} show that different aspects in the data preprocessing and model training could affect the PTM-based AWI approach.
% Such findings motivate researches to design refined techniques (e.g., the elaborated prompts for the PTM-based AWI approach, the customized AWI pre-training tasks, or the meticulous AWI fine-tuning designs for PTMs) to further improve the PTM-based AWI approach. 
% it is very promising to incorporate AWI-specific designs (e.g., fine-grained static analysis information \cite{staticqq} and dynamic execution information \cite{dynamic}) to fine-tuning PTMs.

% However, there is some irrelevant information in the source code from the method containing a warning \cite{staticqq}, which hinders the performance of PTM-based AWI approach.
% Fortunately, 

\subsection{Practical Guidelines}
% Sections \ref{rq1}$\sim$\ref{rq3} show that the studied five PTMs on AWI consistently outperform the SOTA ML-based AWI approach. 
% However, the optimal AUC of the PTM-based AWI approach is only about 70\%. Also, Section \ref{rq4} reveals that the median AUC of the PTM-based AWI approach is nearly 60\% in the within project AWI scenario while only being close to 50\% in the cross project AWI scenario.
% It indicates that although PTMs manifest impressive advancements in AWI, there is still substantial room for improvement. 
% data preprocessing, model training, and model prediction in Section \ref{overview}.
Based on our findings in Section \ref{result}, we highlight practical guidelines for the future AWI community from the perspectives of a typical PTM-based AWI workflow.

\textbf{Incorporating the refined warning context into PTM-based AWI.} Sections \ref{rq1}$\sim$\ref{rq3} show that by simply extracting source code from the method containing a warning as the warning context, PTMs have already achieved a new breakthrough in the AWI field.  
However, the results of Section \ref{rq5} indicate that the warning contexts extracted by our study are still coarse-grained, causing PTM's underperformance on AWI.
In the future, it could be essential to extract the fine-grained warning contexts via the well-designed static analysis techniques (e.g., program slice) \cite{staticqq} and the rigorous dynamic execution tactics (e.g., fuzzing) \cite{dynamic}, thereby capturing the discriminative patterns between actionable and unactionable warnings for PTM-based AWI. 
Also, warnings reported by SCAs have various characteristics (e.g., category and message), which could provide hints for AWI. 
Thus, it could be promising to enrich the warning context with warning characteristics, thereby amplifying PTM-based AWI performance.

% amplify the PTM-based AWI approach by equipping the context with richer warning information.

% In addition to source code from the method containing a warning, warnings reported by SCAs have various attributes (e.g., category, priority, and message) to depict their characteristics. 
% Besides, the fine-grained static analysis information \cite{staticqq} or the refined program dynamic execution data \cite{dynamic} can provide precise hints to identify targeted warnings into actionable and unactionable ones. 
% Thus, it is a promising way to amplify the PTM-based AWI approach by equipping the context with richer warning information.

\textbf{Enlarging the benefits of pre-training and fine-tuning for PTM-based AWI.} Section \ref{rq3} describes that the pre-training and fine-tuning components in PTMs benefit AWI. 
Such findings inspire researchers to explore more innovative AWI approaches from the following aspects.
As for the pre-training component, it is a naive way to apply various PTMs with more abundant and larger code-related pre-training corpora (e.g., CodeLlama \cite{codellama}) for AWI. Besides, it is engaging to develop domain-specific PTMs by formulating pre-training tasks related to AWI. 
As for the fine-tuning component, Fig. \ref{fig:rq32} shows that the upward trend of AUC has not even diminished when all warnings in the training set are included, indicating that AWI performance could be further improved with the additional fine-tuning warning samples. 
In the future, it is necessary to enhance AWI performance by fine-tuning PTMs on more warning datasets.

\textbf{Integrating adequate and diverse warnings with PTM-based AWI.} As concluded in Section \ref{rq4}, increasing the number of training set improves PTM-based AWI performance in within project AWI scenario. 
It indicates that the adequacy of warnings in the training set is important for the PTM-based AWI approach. 
However, due to the heterogeneity of warnings in cross project AWI scenario, increasing the number of training set brings little performance improvement in PTM-based AWI. 
It signifies that the diversity of warnings in the training set is vital to the PTM-based AWI approach. 
In addition, Section \ref{rq5} explicitly states that the adequacy and diversity of warnings in the training set would affect the PTM-based AWI performance. 
In the future, it may be a potential way to gather adequate and diverse warnings in the training set, thereby boosting PTM-based AWI performance.

% \textbf{Enlarging the benefits of fine-tuning for AWI.} Section \ref{rq32} signifies that PTMs with an increased array of fine-tuning corpora basically boost AWI performance. 
% Fig. \ref{fig:rq32} shows that there is an approximately linear relationship between AUC and fine-tuning data size. 
% The upward trend of AUC has not even diminished when all warnings in the training set are included, indicating that AWI performance could be further improved with the additional fine-tuning warning samples. 
% Besides, the fine-tuning is a straightforward and simple way \cite{zhang2023pre}.
% In the future, it is necessary to explore the performance of the PTM-based AWI approach with more warning datasets.

% Further, it is very promising to incorporate AWI-specific designs (e.g., fine-grained static analysis information \cite{staticqq} and dynamic execution information \cite{dynamic}) to fine-tuning PTMs.

% \textbf{Embodying the advance of PTMs for AWI.} In Section \ref{rq1} show that the studied five PTMs on AWI perform better than the state-of-the-art ML-based AWI approach. 
% Besides, Section \ref{rq3} describes that the pre-training and fine-tuning components in PTMs are beneficial for AWI.
% Such a finding inspires us to explore more innovative AWI approaches for by utilizing various PTMs, e.g., developing domain-specific PTMs by formulating pre-training tasks related to AWI.

\subsection{Threats to validity} \label{threat} 
\textbf{External.} Our findings may not be generalized to other PTMs and SCAs. 
To alleviate this threat, as for PTMs, we select five representative PTMs for evaluation. Such PTMs not only show powerful performance in recent code-related tasks \cite{niu2023empirical}, but also cover the three typical PTM architectures. 
As for SCAs, we select SpotBugs for evaluation because SpotBugs (1) spans plentiful bug patterns (i.e., 400+), which is similar to other SCAs (e.g., Infer \cite{Infer}), (2) has been widely used in open-source/commercial projects \cite{useful3}, and (3) is often considered as a typical target in academic research \cite{benchnew11, use1}.
We acknowledge, however, that there are some differences among PTMs and SCAs.
In future work, we will conduct more experiments in other PTMs and SCAs.

\textbf{Internal.} The internal threat to validity is related to the baseline selection in RQ1. 
The existing studies \cite{empiricalex, empirical, S44} experimentally prove the effectiveness of the DL-base AWI approach over the traditional ML-based AWI approach. Thus, we elaborately select a DL-based AWI approach as the SOTA baseline for comparison. 
Besides, we follow the approach design and model selection of these studies to implement the baseline. 
Thus, we believe that such a scrupulous experiment design can mitigate the above threat. 

\textbf{Construct.} The construct threat to validity is related to the warning dataset. We collect the warning dataset from 10 large-scale and open-source projects with various domains and sufficient maturity.
Besides, after using a SOTA multi-stage matching technique \cite{mining} to assign labels for warnings, we conduct the manual inspection and the automatic filter to further ensure the warning label reliability. 
Although there may still be a few noisy warnings in the dataset, these noisy warnings could facilitate the robustness of PTM-based AWI \cite{xu2021differential}.
Also, we plan to collect more ground-truth warning datasets for future PTM-based AWI research.
% However, the of such a technique could be bring mislabeled warnings.
% In future, we plan to optimize the warning labeling quality by enhancing the warning matching technique. 
% Also, we plan to explore the actual advantage of PTM-based AWI in more warning datasets.
\section{Conclusion} \label{conclusion}
In this paper, we conduct the first study to explore the feasibility of PTMs on AWI.
By evaluating 10K+ SpotBugs warnings, the results show that PTMs substantially improve AWI compared to the SOTA ML-based AWI approach. 
Besides, we investigate the impact of data preprocessing ways, model training components, and model detection scenarios in the typical PTM-based AWI workflow.
Moreover, we analyze the incorrect predictions of current PTMs on AWI.
Based on our findings, we further supply practical guidelines for future PTM-based AWI research.
Overall, our study augments the usability of SCAs and exhibits the promising future of using PTMs for AWI. 

% so as to understand the pros and cons of PTMs on AWI. 

\bibliographystyle{IEEEtran}
\bibliography{ref}

\begin{thebibliography}{10}
\providecommand{\url}[1]{#1}
\csname url@samestyle\endcsname
\providecommand{\newblock}{\relax}
\providecommand{\bibinfo}[2]{#2}
\providecommand{\BIBentrySTDinterwordspacing}{\spaceskip=0pt\relax}
\providecommand{\BIBentryALTinterwordstretchfactor}{4}
\providecommand{\BIBentryALTinterwordspacing}{\spaceskip=\fontdimen2\font plus
\BIBentryALTinterwordstretchfactor\fontdimen3\font minus \fontdimen4\font\relax}
\providecommand{\BIBforeignlanguage}[2]{{%
\expandafter\ifx\csname l@#1\endcsname\relax
\typeout{** WARNING: IEEEtran.bst: No hyphenation pattern has been}%
\typeout{** loaded for the language `#1'. Using the pattern for}%
\typeout{** the default language instead.}%
\else
\language=\csname l@#1\endcsname
\fi
#2}}
\providecommand{\BIBdecl}{\relax}
\BIBdecl

\bibitem{static}
B.~Chess and G.~McGraw, ``Static analysis for security,'' \emph{IEEE Symposium on Security and Privacy (S\&P)}, vol.~2, pp. 76--79, 2004.

\bibitem{useful3}
A.~Habib and M.~Pradel, ``How many of all bugs do we find? a study of static bug detectors,'' in \emph{Proceedings of the 33rd ACM/IEEE International Conference on Automated Software Engineering (ASE)}, 2018, pp. 317--328.

\bibitem{useful5}
S.~Lipp, S.~Banescu, and A.~Pretschner, ``An empirical study on the effectiveness of static c code analyzers for vulnerability detection,'' in \emph{Proceedings of the 31st ACM SIGSOFT International Symposium on Software Testing and Analysis (ISSTA)}, 2022, pp. 1--13.

\bibitem{rebalance}
X.~Ge, C.~Fang, T.~Bai, J.~Liu, and Z.~Zhao, ``An empirical study of class rebalancing methods for actionable warning identification,'' \emph{IEEE Transactions on Reliability (TR)}, vol.~72, no.~4, pp. 1--15, 2023.

\bibitem{featureselection}
G.~Xiuting, F.~Chunrong, L.~Jia, Q.~Mingshuang, L.~Xuanye, and Z.~Zhihong, ``An unsupervised feature selection approach for actionable warning identification,'' \emph{Expert Systems with Applications (ESWA)}, vol. 227, p. 120152, 2023.

\bibitem{shab}
R.~Yedida, H.~J. Kang, H.~Tu, X.~Yang, D.~Lo, and T.~Menzies, ``How to find actionable static analysis warnings: A case study with findbugs,'' \emph{IEEE Transactions on Software Engineering (TSE)}, pp. 1--17, 2023.

\bibitem{datasetheihei}
H.~J. Kang, K.~L. Aw, and D.~Lo, ``Detecting false alarms from automatic static analysis tools: how far are we?'' in \emph{Proceedings of the 44th IEEE/ACM International Conference on Software Engineering (ICSE)}, 2022, pp. 698--709.

\bibitem{andreasen2017systematic}
E.~S. Andreasen, A.~M{\o}ller, and B.~B. Nielsen, ``Systematic approaches for increasing soundness and precision of static analyzers,'' in \emph{Proceedings of the 6th ACM SIGPLAN International Workshop on State of the Art in Program Analysis}, 2017, pp. 31--36.

\bibitem{rice}
H.~G. Rice, ``Classes of recursively enumerable sets and their decision problems,'' \emph{Journal of Symbolic Logic}, vol.~74, no.~2, pp. 358--366, 1953.

\bibitem{S44}
A.~Kharkar, R.~Z. Moghaddam, M.~Jin, X.~Liu, X.~Shi, C.~Clement, and N.~Sundaresan, ``Learning to reduce false positives in analytic bug detectors,'' in \emph{Proceedings of the 44th International Conference on Software Engineering (ICSE)}, 2022, pp. 1307--1316.

\bibitem{bielik2017learning}
P.~Bielik, V.~Raychev, and M.~Vechev, ``Learning a static analyzer from data,'' in \emph{Proceedings of the 29th International Conference of Computer Aided Verification (CAV)}, 2017, pp. 233--253.

\bibitem{survey1}
T.~Muske and A.~Serebrenik, ``Survey of approaches for postprocessing of static analysis alarms,'' \emph{ACM Computing Survey (CSUR)}, vol.~55, no.~3, 2022.

\bibitem{ge2023machine}
X.~Ge, C.~Fang, X.~Li, W.~Sun, D.~Wu, J.~Zhai, S.~Lin, Z.~Zhao, Y.~Liu, and Z.~Chen, ``Machine learning for actionable warning identification: A comprehensive survey,'' \emph{arXiv preprint arXiv:2312.00324}, 2023.

\bibitem{empirical}
U.~Koc, S.~Wei, J.~S. Foster, M.~Carpuat, and A.~A. Porter, ``An empirical assessment of machine learning approaches for triaging reports of a java static analysis tool,'' in \emph{Proceedings of the 12th IEEE Conference on Software Testing, Validation and Verification (ICST)}, 2019, pp. 288--299.

\bibitem{empiricalex}
S.~Yerramreddy, A.~Mordahl, U.~Koc, S.~Wei, J.~S. Foster, M.~Carpuat, and A.~A. Porter, ``An empirical assessment of machine learning approaches for triaging reports of static analysis tools,'' \emph{Empirical Software Engineering (EMSE)}, vol.~28, no.~2, p.~28, 2023.

\bibitem{sanxing}
S.~Lee, S.~Hong, J.~Yi, T.~Kim, C.-J. Kim, and S.~Yoo, ``Classifying false positive static checker alarms in continuous integration using convolutional neural networks,'' in \emph{Proceedings of the 12th IEEE Conference on Software Testing, Validation and Verification (ICST)}, 2019, pp. 391--401.

\bibitem{MAPL}
U.~Koc, P.~Saadatpanah, J.~S. Foster, and A.~A. Porter, ``Learning a classifier for false positive error reports emitted by static code analysis tools,'' in \emph{Proceedings of the 1st ACM SIGPLAN International Workshop on Machine Learning and Programming Languages (MAPL)}, 2017, pp. 35--42.

\bibitem{smartcontract}
K.~T. Tran and H.~D. Vo, ``Scar: smart contract alarm ranking,'' in \emph{Proceedings of the 29th Asia-Pacific Software Engineering Conference (APSEC)}, 2022, pp. 447--451.

\bibitem{pretrainedmodel}
X.~Han, Z.~Zhang, N.~Ding, Y.~Gu, X.~Liu, Y.~Huo, J.~Qiu, Y.~Yao, A.~Zhang, L.~Zhang \emph{et~al.}, ``Pre-trained models: Past, present and future,'' \emph{AI Open}, vol.~2, pp. 225--250, 2021.

\bibitem{zhang2023pre}
Q.~Zhang, C.~Fang, B.~Yu, W.~Sun, T.~Zhang, and Z.~Chen, ``Pre-trained model-based automated software vulnerability repair: How far are we?'' \emph{IEEE Transactions on Dependable and Secure Computing (TDSC)}, pp. 1--18, 2023.

\bibitem{spotbigs}
SpotBugs, \url{https://spotbugs.github.io/.}, lasted accessed March 10, 2024.

\bibitem{github}
R.~package, \url{https://sites.google.com/view/ptm4awidata}.

\bibitem{ViolationTracker}
P.~Yu, Y.~Wu, X.~Peng, H.~Peng, J.~Zhang, P.~Xie, and W.~Zhao, ``Violationtracker: Building precise histories for static analysis violations,'' in \emph{Proceedings of the 45th International Conference on Software Engineering (ICSE)}, 2023, pp. 1--12.

\bibitem{mining}
K.~Liu, D.~Kim, T.~F. Bissyandé, S.~Yoo, and Y.~Le~Traon, ``Mining fix patterns for findbugs violations,'' \emph{IEEE Transactions on Software Engineering (TSE)}, vol.~47, no.~1, pp. 165--188, 2021.

\bibitem{golden}
J.~Wang, S.~Wang, and Q.~Wang, ``Is there a ``golden'' feature set for static warning identification? an experimental evaluation,'' in \emph{Proceedings of the 12th ACM/IEEE International Symposium on Empirical Software Engineering and Measurement (ESEM)}, 2018, pp. 1--10.

\bibitem{limit}
J.~Wang, Y.~Huang, S.~Wang, and Q.~Wang, ``Find bugs in static bug finders,'' in \emph{Proceedings of the 30th International Conference on Program Comprehension (ICPC)}, 2022, pp. 516--527.

\bibitem{llmclassification}
F.~F. Xu, U.~Alon, G.~Neubig, and V.~J. Hellendoorn, ``A systematic evaluation of large language models of code,'' in \emph{Proceedings of the 6th ACM SIGPLAN International Symposium on Machine Programming (MAPS)}, 2022, p. 1–10.

\bibitem{vaswani2017attention}
A.~Vaswani, N.~Shazeer, N.~Parmar, J.~Uszkoreit, L.~Jones, A.~N. Gomez, {\L}.~Kaiser, and I.~Polosukhin, ``Attention is all you need,'' \emph{Advances in neural information processing systems}, vol.~30, 2017.

\bibitem{BRADLEY19971145}
A.~P. Bradley, ``The use of the area under the roc curve in the evaluation of machine learning algorithms,'' \emph{Pattern Recognition}, vol.~30, no.~7, pp. 1145--1159, 1997.

\bibitem{ge2021impact}
X.~Ge, Y.~Huang, Z.~Hui, X.~Wang, and X.~Cao, ``Impact of datasets on machine learning based methods in android malware detection: an empirical study,'' in \emph{Proceedings of the 21st International Conference on Software Quality, Reliability and Security (QRS)}, 2021, pp. 81--92.

\bibitem{a8}
X.~Yang, Z.~Yu, J.~Wang, and T.~Menzies, ``Understanding static code warnings: An incremental ai approach,'' \emph{Expert System with Applications (ESWA)}, vol. 167, p. 114134, 2021.

\bibitem{a10}
X.~Yang, J.~Chen, R.~Yedida, Z.~Yu, and T.~Menzies, ``Learning to recognize actionable static code warnings (is intrinsically easy),'' \emph{Empirical Software Engineering (EMSE)}, vol.~26, no.~3, 2021.

\bibitem{codex}
M.~Chen, J.~Tworek, H.~Jun, Q.~Yuan, H.~P. d.~O. Pinto, J.~Kaplan, H.~Edwards, Y.~Burda, N.~Joseph, G.~Brockman \emph{et~al.}, ``Evaluating large language models trained on code,'' \emph{arXiv preprint arXiv:2107.03374}, 2021.

\bibitem{brown2020language}
T.~Brown, B.~Mann, N.~Ryder, M.~Subbiah, J.~D. Kaplan, P.~Dhariwal, A.~Neelakantan, P.~Shyam, G.~Sastry, A.~Askell \emph{et~al.}, ``Language models are few-shot learners,'' \emph{Advances in Neural Information Processing Systems (NIPS)}, vol.~33, pp. 1877--1901, 2020.

\bibitem{raffel2020exploring}
C.~Raffel, N.~Shazeer, A.~Roberts, K.~Lee, S.~Narang, M.~Matena, Y.~Zhou, W.~Li, and P.~J. Liu, ``Exploring the limits of transfer learning with a unified text-to-text transformer,'' \emph{The Journal of Machine Learning Research}, vol.~21, no.~1, pp. 5485--5551, 2020.

\bibitem{radford2019language}
A.~Radford, J.~Wu, R.~Child, D.~Luan, D.~Amodei, I.~Sutskever \emph{et~al.}, ``Language models are unsupervised multitask learners,'' \emph{OpenAI blog}, vol.~1, no.~8, p.~9, 2019.

\bibitem{niu2023empirical}
C.~Niu, C.~Li, V.~Ng, D.~Chen, J.~Ge, and B.~Luo, ``An empirical comparison of pre-trained models of source code,'' in \emph{Proceedings of the 45th International Conference on Software Engineering (ICSE)}, 2023, pp. 2136--2148.

\bibitem{huggingface}
H.~Face, \url{https://huggingface.co/.}, lasted accessed February 18, 2024.

\bibitem{codebert}
Z.~Feng, D.~Guo, D.~Tang, N.~Duan, X.~Feng, M.~Gong, L.~Shou, B.~Qin, T.~Liu, D.~Jiang \emph{et~al.}, ``Codebert: A pre-trained model for programming and natural languages,'' \emph{Findings of the Association for Computational Linguistics (ACL)}, 2020.

\bibitem{guo2020graphcodebert}
D.~Guo, S.~Ren, S.~Lu, Z.~Feng, D.~Tang, S.~Liu, L.~Zhou, N.~Duan, A.~Svyatkovskiy, S.~Fu \emph{et~al.}, ``Graphcodebert: Pre-training code representations with data flow,'' \emph{arXiv preprint arXiv:2009.08366}, 2020.

\bibitem{wang2021codet5}
Y.~Wang, W.~Wang, S.~Joty, and S.~C. Hoi, ``Codet5: Identifier-aware unified pre-trained encoder-decoder models for code understanding and generation,'' in \emph{Proceedings of the 26th International Conference on Empirical Methods in Natural Language Processing (EMNLP)}, 2021.

\bibitem{guo2022unixcoder}
D.~Guo, S.~Lu, N.~Duan, Y.~Wang, M.~Zhou, and J.~Yin, ``Unixcoder: Unified cross-modal pre-training for code representation,'' in \emph{Proceedings of the 60th Annual Meeting of the Association for Computational Linguistics (ACL)}, 2022.

\bibitem{lu2021codexglue}
S.~Lu, D.~Guo, S.~Ren, J.~Huang, A.~Svyatkovskiy, A.~Blanco, C.~Clement, D.~Drain, D.~Jiang, D.~Tang \emph{et~al.}, ``Codexglue: A machine learning benchmark dataset for code understanding and generation,'' \emph{arXiv preprint arXiv:2102.04664}, 2021.

\bibitem{sdp-delay}
G.~G. Cabral, L.~L. Minku, E.~Shihab, and S.~Mujahid, ``Class imbalance evolution and verification latency in just-in-time software defect prediction,'' in \emph{Proceedings of the 41st IEEE/ACM International Conference on Software Engineering (ICSE)}, 2019, pp. 666--676.

\bibitem{observe}
S.~Kim and M.~D. Ernst, ``Prioritizing warning categories by analyzing software history,'' in \emph{Proceedings of the 4th International Workshop on Mining Software Repositories (MSR)}, 2007, pp. 27--27.

\bibitem{pytorch}
PyTorch, \url{https://pytorch.org/.}, lasted accessed March 10, 2024.

\bibitem{sgd}
Y.~Liu, Y.~Gao, and W.~Yin, ``An improved analysis of stochastic gradient descent with momentum,'' in \emph{Proceedings of the 34th International Conference on Neural Information Processing Systems (NIPS)}, 2020.

\bibitem{kingma2014adam}
D.~P. Kingma and J.~Ba, ``Adam: A method for stochastic optimization,'' 2015, pp. 1--15.

\bibitem{vrepair}
Z.~Chen, S.~Kommrusch, and M.~Monperrus, ``Neural transfer learning for repairing security vulnerabilities in c code,'' \emph{IEEE Transactions on Software Engineering (TSE)}, 2022.

\bibitem{gra2}
S.~Kim, S.~Woo, H.~Lee, and H.~Oh, ``Vuddy: A scalable approach for vulnerable code clone discovery,'' in \emph{2017 IEEE Symposium on Security and Privacy (S\&P)}, 2017, pp. 595--614.

\bibitem{javaparse}
JavaParser, \url{https://github.com/javaparser/javaparser.}, lasted accessed February 18, 2024.

\bibitem{krawczyk2016learning}
B.~Krawczyk, ``Learning from imbalanced data: open challenges and future directions,'' \emph{Progress in Artificial Intelligence}, vol.~5, no.~4, pp. 221--232, 2016.

\bibitem{staticqq}
T.~Muske, R.~Talluri, and A.~Serebrenik, ``Repositioning of static analysis alarms,'' in \emph{Proceedings of the 27th ACM SIGSOFT International Symposium on Software Testing and Analysis (ISSTA)}, 2018, pp. 187--197.

\bibitem{dynamic}
A.~Kallingal~Joshy, X.~Chen, B.~Steenhoek, and W.~Le, ``Validating static warnings via testing code fragments,'' in \emph{Proceedings of the 30th ACM SIGSOFT International Symposium on Software Testing and Analysis (ISSTA)}, 2021, pp. 540--552.

\bibitem{codellama}
B.~Rozière, J.~Gehring, F.~Gloeckle, S.~Sootla, I.~Gat, X.~Tan, Y.~Adi, J.~Liu, T.~Remez, J.~Rapin, A.~Kozhevnikov, I.~Evtimov, J.~Bitton, M.~Bhatt, C.~Ferrer, A.~Grattafiori, W.~Xiong, A.~Defossez, J.~Copet, and G.~Synnaeve, ``Code llama: Open foundation models for code,'' in \emph{arXiv:2308.12950}, 2023.

\bibitem{Infer}
Infer, \url{https://fbinfer.com/.}, lasted accessed February 18, 2024.

\bibitem{benchnew11}
J.~Li, ``A better approach to track the evolution of static code warnings,'' in \emph{Proceedings of the 43rd International Conference on Software Engineering: Companion Proceedings (ICSE-Companion)}, 2021, pp. 135--137.

\bibitem{use1}
D.~Marcilio, C.~A. Furia, R.~Bonif{\'{a}}cio, and G.~Pinto, ``Spongebugs: Automatically generating fix suggestions in response to static code analysis warnings,'' \emph{Journal of Systems and Software (JSS)}, vol. 168, p. 110671, 2020.

\bibitem{xu2021differential}
J.~Xu, Y.~Li, and R.~H. Deng, ``Differential training: A generic framework to reduce label noises for android malware detection,'' in \emph{Proceedings of the 28th Network and Distributed Systems Security Symposium (NDSS)}, 2021, pp. 1--14.

\end{thebibliography}

\end{document}